\documentclass{revtex4-2}

\usepackage[utf8]{inputenc}
\usepackage[english]{babel}

\usepackage[hidelinks]{hyperref}

\usepackage{graphicx}
\usepackage{subcaption}
\usepackage{parskip}

\usepackage{float}
\usepackage{booktabs}
\usepackage{makecell}
\usepackage{multirow}

\usepackage[figurename=Fig.]{caption}
\usepackage[tablename=Tab.]{caption}
\usepackage[font=small]{caption}

\usepackage{amsmath}
\usepackage{amssymb}
\usepackage{amsfonts}
\usepackage{mathtools}

\usepackage{upgreek}
\usepackage{xcolor}
\usepackage[normalem]{ulem}

\begin{document}


\title{Comparison of Magnetized Thick Disks around Black Holes and Boson Stars}
\date{\today}

\author{Kristian Gjorgjieski}
 \email{kristian.gjorgjieski@uol.de}
\author{Jutta Kunz}%
 \email{jutta.kunz@uni-oldenburg.de}
\affiliation{Department of Physics, Carl von Ossietzky University of Oldenburg, 26111 Oldenburg, Germany}

\author{Petya Nedkova}
 \email{pnedkova@phys.uni-sofia.bg}
\affiliation{Department of Theoretical Physics, Sofia University, Sofia 1164, Bulgaria}

\begin{abstract}
Boson stars are considered as promising candidates for black hole mimickers. Similar to other compact objects they can form accretion disks around them. The properties of these disks could possibly distinguish them from other compact objects like black holes in future observations. Retrograde thick disks around boson stars and the influence of strong magnetic fields on them were already studied and it was shown that they can harbor very distinct features compared to black hole disks. However, the case of prograde thick disks is mostly unexplored, since they may appear much more similar to black hole disks. In this work we will investigate similarities and differences regarding prograde thick disks around non-selfinteracting rotating boson stars and rotating black holes. We assume thereby a polytropic equation of state and a constant specific angular momentum distribution of the disks. We classify the various conceivable boson star and black hole solutions by a dimensionless spin parameter $a$ and compare their corresponding disk solutions. The influence of toroidal magnetic fields on the disks is analyzed by selected disk properties, as the rest-mass density distribution and the Bernoulli parameter. Disk solutions are characterized by their degree of magnetization represented by the magnetization parameter $\beta_{mc}$. We found that strong magnetic fields can strengthen the differences of disk solutions or oppositely even lead to a correlation in disk properties, depending on the spin parameter of the boson star and black hole solutions. We identify the vertical thickness of the boson star disks as the main differentiating factor, since for most solutions the vertical density distribution is far more outreaching for boson star disks compared to black hole disks.
\end{abstract}

\maketitle


\section{Introduction}
Accretion disks are objects composed of gas and plasma particles orbiting around compact objects. They form when attracted matter takes stable orbits around a central object, where then further phenomena like friction and magnetohydrodynamical effects induce instabilities, leading to an inwards spiral of matter into the central object. The acting gravitational forces are compressing the disk matter and the resulting friction is heating the disk up. The generated heat is further transformed into radiation energy and leaves the disk in form of electromagnetic radiation, whose frequency range depends on the properties of the disk and the central object. Since the energy emitted by the disk scales with the gravitational field, accretion disks around the centers of galaxies are among the brightest objects observed in the universe and are therefore also a topic of interest in the study of high energy physics.

In general the properties of accretion disks depend on a complex regime of magnetohydrodynamical and gravitational effects \cite{FoundationsADTheory,Czerny_2023,Lasota_2016,Abramowicz:1971, hou2024new}. One of the main crucial aspects in their formation is the acting gravitational field and therefore the local spacetime background, which depends on the central object, since in most cases the self-gravitational interaction of the disk is small compared to the gravitational interaction with the central object and therefore negligible. Thus, the study and observation of accretion disks is closely related to the study of the compact objects forming them \cite{EBISAWA20062862,white_1997,Fabrika_2015,C_rdenas_Avenda_o_2019,Maccarone_2013,speri2022measuring}. Due to the recent progress in the observation of distant objects, especially the progress of the Event Horizon telescope \cite{EHT_1, EHT_10, EHT_2}, it is expected that a deeper observational investigation of accretion disks will be possible in the near future. This will eventually allow us to learn more about the compact objects themselves, around which they form, and about the high-energy processes occurring inside the disks.  

A broad spectrum of astrophysical objects are capable of forming accretion disks around them, ranging from protostars to neutron stars and from stellar black holes to supermassive galactic nuclei.
In theory, there exist many more hypothetical objects, yet to be actually observed, which are capable of forming accretion disks around them. Some of these objects are described by theoretical predictions in such a way, that they could appear similar to black holes for a distant observer. One promising candidate for such black hole mimickers are boson stars. For a distant observer they could resemble black holes in particular due to their negligible radiation profile and their strong gravitational field. Boson stars originate from the works of John A. Wheeler about \textit{geons} \cite{Geons}, where he developed the idea of self-gravitating electromagnetic fields. These ideas where further developed by considering a complex scalar field, minimally coupled to gravity, instead of electromagnetic fields \cite{Klein-Gordon-Geon,ScalarParticle,Self-Gravitating-Particles}. This approach led to stable solutions, which are now known as boson stars. In general they are composed of spin-0 particles in the form of compact scalar field condensates, which form an equilibrium state after a self-gravitational collapse of the massive scalar field. The counteracting force originates from the uncertainty principle, which prevents further gravitational collapse and stabilises the boson star \cite{Jetzer, Schunck, Liebling}. Depending on the model (i.e. described by the chosen potential and the chosen symmetries), boson stars may have different properties and a broad range of various configurations have been already described and analyzed \cite{Kaup, Feinblum, Ruffini, Kobayashi, Yoshida, Schunck_1998, Kleihaus, Kleihaus_2008, Collodel_2017, Collodel_2019, Rosa_2022, Rosa_2022_observational_signatures, Rosa_2023}. Their size and mass can range from the atomic scale to the galactic scale. However, it should be mentioned that they do not have a sharp surface, but exhibit an exponential decay asymptotically. To determine their mass and radius, a definition by compactness is usually chosen. Due to their negligible radiation profile, they are not only potential black hole mimickers, but also promising candidates for dark matter \cite{BS_dark_matter,BS_DMclumps}.
One way to detect them would be by indirect observation of the accretion disks that could form around them, in case that one could distinguish these accretion disks from those that would form around other objects like black holes. In that regard, they may possess bound circular orbits which also extend into the inner regions with high density of the scalar field distribution in contrast to black holes \cite{Meliani}.
Additionally, the formation of a shadow in the strict sense would be prevented since they are horizonless, but nevertheless (in the presence of accretion disks) a central dark region will be observed similar to that of black holes, which corresponds to the lensed image of an accretion disk's inner edge \cite{Vincent_2016, Olivares, Vincent_2021, EHT_Sgr}.

In this work we will compare magnetized thick accretion disks of rotating uncharged black holes with those of rotating boson stars without scalar field self-interaction. The thick disk model offers a simple but realistic analytical description of a (non-selfgravitating) accretion disk \cite{Fishbone1,Abramowicz, Kozlowski}. In this model the accretion disk is modelled as perfect fluid orbiting the compact object and the fluid particles obey a polytropic equation of state, where the pressure is a function of the rest-mass density. The simplest form of thick disks solutions are called \textit{Polish doughnuts}, where the electromagnetic, viscosity and radiation terms are neglected and only the fluid part enters the equations. The resulting disk solutions can be examined by their corresponding effective potential, which is linked through the relativistic Euler equations to thermodynamic quantities and offers therefore a general description of the disk. This effective potential depends only on the metric coefficients and the specific angular momentum of the fluid particles. The thick disk model represents therefore a simple model for making qualitative statements about the accretion disks and the accretion process and is in this respect a suitable complement to computationally complex numerical methods. Over the last decades the properties of thick disks and thick disk accretion onto black holes were studied in various works \cite{Abramowicz_1983,Font_2002,ThickDiskOcillations,TiltedThickDisks, Cassing:2023}. The purely hydrodynamical Polish doughnut model can be extended by adding a toroidal magnetic field to obtain an astrophysically more accurate model \cite{Komissarov:2006nz,Font,Wielgus,Montero}, since it is widely believed that the dynamics of astrophysical accretion disks are strongly influenced by magnetic fields. We will use this magnetized model and furthermore we will assume a constant specific angular momentum distribution of the disk, where the specific angular momentum enters the equations as a free parameter. Solutions with a constant specific angular momentum play a fundamental role, since despite their simplicity they already capture the main qualitative features of more complex configurations. The disk solutions will be classified by a magnetization parameter $\beta_{mc}$, in order to distinguish between the effects of different degrees of magnetization on the disk properties.

Retrograde Polish doughnuts around boson stars without self-interaction have already been studied in \cite{Teodoro_2021b}, and it was shown that they can harbour unique features like two-centered disk solutions and static surfaces. Magnetic fields can greatly influence these properties and alter the unique features of retrograde thick disks as it was shown in \cite{Gjorgjieski}. Since these retrograde disks can already be very distinct compared to retrograde magnetized thick disks around black holes, there is less of a need for a deeper comparison between them (we will address them briefly in a later section). Therefore our focus will be on prograde disks, which appear in principle much more similar to black hole disks and which have not been examined deeply so far. In particular we will conduct a qualitative comparison between boson star and black hole disks, where we try to isolate the influence of the magnetization as much as possible, by comparing boson star and black hole solutions with the same spin parameter $a$ and the same specific angular momentum $\ell_0$ of their disks.

In the following section we will describe the black hole and boson star solutions which we will consider, afterwards section 3 reviews the theory of magnetized thick disks and their construction procedure. Section 4 presents the disk comparison and section 5 briefly focuses on retrograde disks. Throughout this work we will use the Einstein summation convention and $(-,+,+,+)$ as the metric signature.

\section{Rotating Boson Stars and Black Holes}
Speaking of astrophysical black holes, it is generally believed that they may possess a non-zero angular momentum and negligible charge. Therefore we consider uncharged rotating black holes (BHs) described by the Kerr metric, which can be written in Boyer-Lindquist coordinates as,
\begin{align}
    ds^2 &= - \left( 1 - \frac{2 M r}{\Sigma} \right) dt^2 - \frac{4 M^2 \Tilde{a} r \sin^2{\theta}}{\Sigma} dt d\varphi + \frac{\Sigma}{\Delta} dr^2 + \Sigma d\theta^2 + \left(r^2 + M^2 \Tilde{a}^2 + \frac{2 M^3 \Tilde{a}^2 r \sin^2{\theta}}{\Sigma} \right) \sin^2{\theta} d\varphi^2 ,\\
    \Delta &\equiv r^2 - 2Mr + M^2\Tilde{a}^2 \ \ ; \ \  \Sigma \equiv r^2 + M^2 \Tilde{a}^2 \cos^2 \theta,
\end{align}
where $M$ is the mass and $\Tilde{a} = \frac{J}{M^2}$ is the dimensionless spin parameter defined by the mass and the angular momentum $J$. Further on we will work with the dimensionless spin parameter, it is therefore convenient to omit the tilde and just use $a$ for the dimensionless spin parameter.
In the case of boson stars (BSs) we consider non-self-interacting BSs obtained from the action $\mathcal{S}$,
\begin{align}
    \mathcal{S} = \int \sqrt{-g} \left(\frac{R}{16 \pi G} - \mathcal{L}_m \right) d^4x,
    \label{eq:action}
\end{align}
where $g$ is the metric determinant, $R$ is the Ricci scalar, $G$ is the gravitational constant and $\mathcal{L}_m = |\partial_\mu \phi |^2 + m^2 |\phi|^2$ is the Lagrangian, with $\phi$ as the complex scalar field and $m$ as the mass of the boson particle. Variation of eq. (\ref{eq:action}) leads to a coupled set of Einstein-Klein-Gordon equations,
\begin{align}
    R_{\mu\nu} - \frac{1}{2} R g_{\mu\nu} &= 8 \pi G T_{\mu\nu}, \\
    \left(\Box - m^2 \right)\phi &= 0,
\end{align}
with $T_{\mu\nu} = (\partial_\mu \phi \partial_\nu \phi^* + \partial_\nu \phi \partial_\mu \phi^*) - \mathcal{L}_m g_{\mu\nu}$ as the stress-energy tensor and $\Box$ as the covariant d'Alembert operator. It should be noted that the action (\ref{eq:action}) is invariant under global U(1) transformations, giving rise to a conserved Noether current $j_\mu$ and Noether charge $Q$,
\begin{align}
j_\mu = i(\phi \partial_\mu \phi^* - \phi^* \partial_\mu \phi) \ \ ; \ \ Q = \int j^t\sqrt{-g}dx^3.
\label{eq:current}
\end{align}
Considering now rotating solutions, the metric should be axisymmetric and stationary, which gives rise to two Killing vector fields, $\partial_t$ and $\partial_\varphi$. The complex scalar field must nevertheless depend on all four coordinates. If the complex scalar field would not depend on the time coordinate, then the conserved charge would be zero (eq.~(\ref{eq:current})) leading to trivial vacuum solutions. Furthermore the stress-energy tensor must depend on the azimuthal coordinate in order to describe rotating solutions. Consequently, the complex scalar field must depend on the time and the azimuthal coordinate in a way that the metric remains unaffected. This can be ensured by decomposing the scalar field into a real profile function and a complex phase,
\begin{align}
    \phi(\mathbf{r},t) = \phi_0(r,\theta) e^{i(\omega t - k \varphi)},
    \label{eq:scalar_field}
\end{align}
where $\phi_0$ is the real valued profile function, $\omega$ is the angular frequency of the scalar field rotation and $k$ is the azimuthal winding number. The azimuthal winding number counts the nodes of the real and imaginary parts of the scalar field along the azimuthal direction and indicates therefore the strength of the angular excitations. Due to single-valuedness of the scalar field, $\phi(\varphi=0)=\phi(\varphi=2\pi)$, $k$ must be an integer. The total angular momentum of the BS is given by the quantisation relation $J = kQ$. In this work we will focus on the fundamental branch of solutions with $k=1$. Making use of the axial symmetry and stationarity, the line element of the metric can be written in the general form as
\begin{align}
    ds^2 = -\alpha^2 dt^2 + A^2(dr^2 + r^2 d \theta^2) + B^2 r^2 \sin^2\theta (d\varphi + \beta^\varphi dt)^2,
\end{align}
where $\alpha$ is the lapse function, $\beta^\varphi$ is the shift function and where the functions $A, B, \alpha$ and $\beta^\varphi$ only depend on $r$ and $\theta$. Substitution of the harmonic ansatz (eq.~(\ref{eq:scalar_field})) into the field equations leads to a coupled set of differential equations which can be solved numerically. Solutions only exist for a set of angular frequencies $\omega$. The solutions are asymptotically flat and the scalar field decays exponentially proportional to $e^{-\sqrt{m^2-\omega^2}r}/r$ asymptotically.
The equations possess a scaling symmetry, which we use to introduce dimensionless quantities by scaling with the boson particle mass $m$, i.e. $\Tilde{r} = rm, \ \Tilde{\omega} = \frac{\omega}{m}$ and $\Tilde{M} = M m$. Further on we will omit the tilde for brevity. The solutions range from the minimum frequency at $\omega = 0.646$ to $\omega = 1$, featuring several branches, with the fundamental branch ranging from the vacuum ($\omega=1$) to the BS with maximum mass.
Solutions below $\omega = 0.665$ possess an ergoregion. Here we will focus only on BS solutions without ergoregions \cite{Ergoregion-instability}.

\begin{figure}[H]
    \centering
    \includegraphics[width=0.8\linewidth]{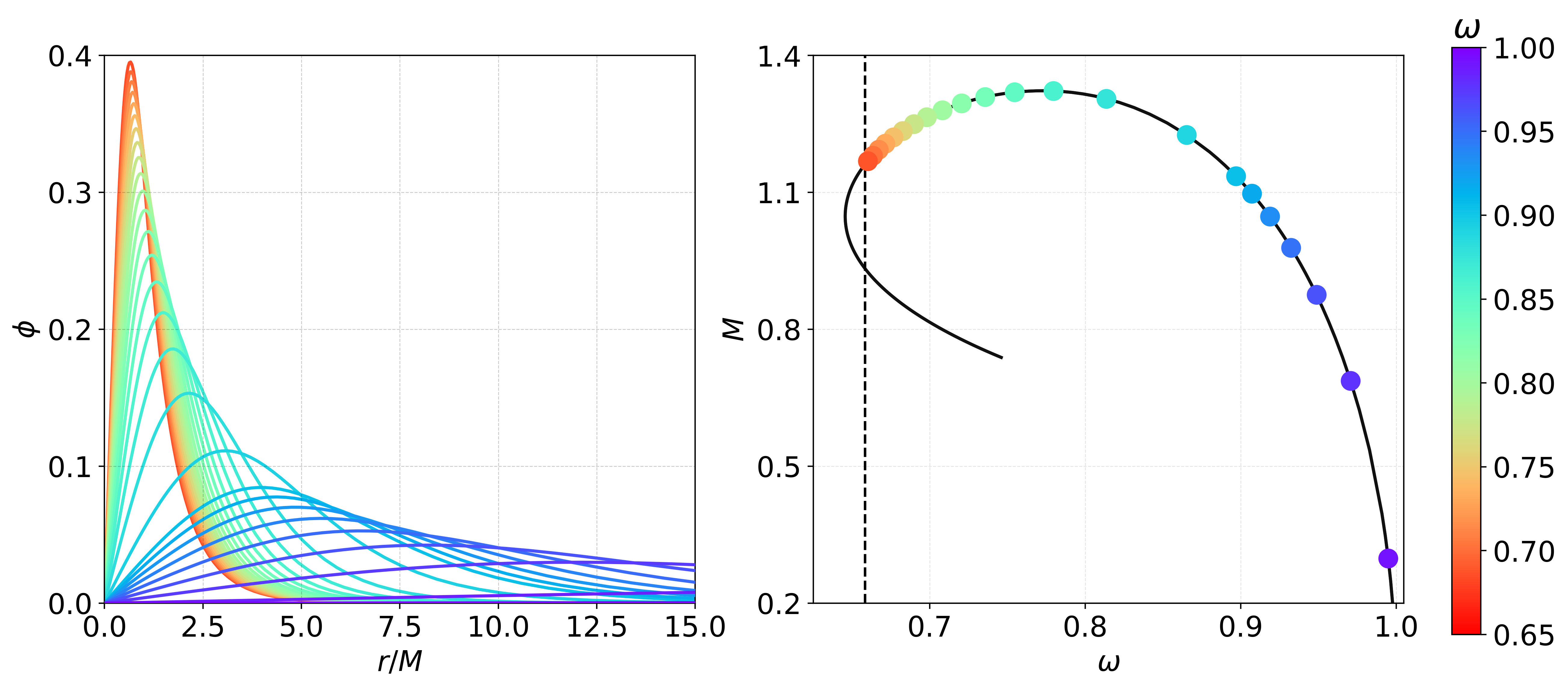}
    \caption{Modulus of the scalar field in the equatorial plane vs.~the scaled radial coordinate on the left side and BS mass vs.~the angular frequency on the right side. Solutions exist only for a set of $\omega$ values, some of which are marked in the plot by circles. The dashed vertical line marks the onset of solutions possessing an ergoregion.}
    \label{fig:BS_solutions}
\end{figure}

The main difference regarding the spacetime environment of rotating BSs and rotating BHs lies in the toroidal shape of the BSs. As seen in Fig.~\ref{fig:BS_solutions} the maximum of the scalar field (and therefore the maximum of the energy-density) is located off-center resulting in a toroidal shape of the scalar field, which moves closer to the center with decreasing $\omega$. Therefore the torus formed by the scalar field is more strongly localized towards the BS center for lower values of $\omega$. 
Due to this toroidal energy density distribution the spacetime near the BS is curved in an unusual way, featuring distinctive orbits as compared to BHs \cite{Collodel_2017a,Teodoro_2021b}.


\section{Magnetized Thick Disks}
Geometrically thick disks can be constructed by linking thermodynamic quantities to kinematic quantities through the relativistic Euler equations \cite{Abramowicz}. In this procedure the disk is modelled as a perfect fluid, where the fluid particles move on circular orbits around a given compact object. To get equilibrium solutions axial symmetry and stationarity of the background spacetime are assumed. Furthermore, as a necessary integrability condition for the relativistic Euler equations, the fluid must obey a barotropic equation of state where the pressure of the fluid can be expressed through its rest-mass density. As a consequence the von Zeipel theorem holds, which states that the angular velocity $\Omega$ and specific angular momentum $\ell$ of a fluid particle can be expressed as a function of each other and the metric components,
\begin{align}
    \ell(r,\theta) = -\frac{\Omega g_{\varphi \varphi} + g_{t\varphi}}{\Omega g_{t\varphi} + g_{tt}} \ \ ; \ \ \Omega(r,\theta) = -\frac{\ell g_{tt} + g_{t\varphi}}{\ell g_{t\varphi} + g_{\varphi \varphi}}.
\end{align}
At the center of the accretion disk the acting forces cancel each other, the flow of the fluid particles at the center follows therefore a geodesic. Their corresponding specific angular momentum $\ell_0$ is then identical to that of a Keplerian circular orbit, $\ell_K$. The general disk properties in the equatorial plane are therefore determined by the equatorial Keplerian specific angular momentum of the spacetime. Assuming a constant specific angular momentum distribution of the disk we can make the following statements: If $\ell_0 \neq \ell_K(r) \forall r$, then no disk solution exists. If $ \exists ! r : \ell_0 = \ell_K(r)$, then a disk solutions exists where the disk center is located at the value of $r$ which satisfies the equation. If there is more than one intersection, the slope of $\ell_K$ determines if it is a center (stable geodesic motion) or a cusp (unstable geodesic motion), the former being given by a positive slope and the latter by a negative. A cusp marks the intersection of equi-pressure (and equi-density) surfaces. We will use the specific angular momentum $\ell_0$ as a free parameter which determines the general solution type (depending on the Keplerian specific angular distribution).

The purely hydrodynamical model can be extended by adding a toroidal magnetic field in order to construct magnetized thick disks. We follow here the procedure described in \cite{Komissarov:2006nz, Font, Gjorgjieski}. The magnetic part enters the stress-energy tensor only additively, and the contravariant stress-energy tensor for the fluid can then be expressed as
\begin{align}
    T^{\mu\nu} &\equiv (\rho h + b^2)u^\mu u^\nu + \left(p_{th} + p_{mag} \right) g^{\mu \nu} - b^\mu b ^\nu,
\end{align}
where $\rho$ is the rest-mass density, $h$ is the specific enthalpy, $u^\mu$ is the four-velocity, $p_{th}$ is the thermodynamic pressure, $p_{mag} = \frac{b^2}{2} \equiv \frac{1}{2} b^\mu b_\mu$ is the magnetic pressure, $g^{\mu\nu}$ is the contravariant metric tensor and $b^\mu = (0, \mathbf{B})$ is the magnetic field with $\mathbf{B}$ being the co-moving three-dimensional magnetic field in the fluid frame.
The relativistic Euler equations can be derived by inserting the stress-energy tensor in the fundamental conservation laws of general relativistic magnetohydrodynamics. The conservation laws are given by,
\begin{align}
    \nabla_\mu (\rho u^\mu) &= 0  \label{eq:continuity}, \\
    \nabla_\mu T^{\mu \nu} &= 0   \label{eq:stress-energy-conserv}, \\
    \nabla_\mu (^*F^{\mu \nu}) &= 0, \label{eq:maxwell}
\end{align}
where $^*F^{\mu \nu} = u^\mu b^\nu - u^\nu b^\mu$ is the dual Faraday tensor. In order to solve this system of equations we assume axial symmetry and stationarity. As a consequence all partial derivatives regarding the time and azimuthal coordinate ($\partial_t$ and $\partial_\varphi$) vanish. Furthermore we assume circular radial motion for the fluid ($u^r = u^\theta = 0$) and make use of the toroidal magnetic field distribution ($b^r = b^\theta = 0$), which leaves the continuity equation (eq.~(\ref{eq:continuity})) and the Maxwell equation (eq.~(\ref{eq:maxwell})) trivially satisfied. To solve eq.~(\ref{eq:stress-energy-conserv}), it can be split into one equation containing the energy conservation and three equations containing the momentum conservation by applying the orthogonal projection tensor $h^\alpha_\beta = \delta^\alpha_\beta + u^\alpha u_\beta$, which leads to
\begin{align}
     (\rho h + b^2)u_\nu \partial_\mu u^\nu + \partial_\mu \left(p_{th} + p_{mag} \right) - b_\nu \partial_\mu b^\nu = 0,
\end{align}
where $\mu = t,\varphi$ are trivially satisfied. Simplifying the equation by rewriting it in terms of the specific angular momentum and the angular velocity of the fluid leads to \cite{Komissarov:2006nz},
\begin{align}
    \partial_\mu(\ln|u_t|) - \frac{\Omega \partial_\mu \ell}{1- \ell \Omega} + \frac{\partial_\mu p_{th}}{\rho h} + \frac{\partial_\mu(\mathcal{L} p_{mag})}{\mathcal{L}\rho h} = 0,
\end{align}
where $\mathcal{L} \equiv g^2_{t\varphi} - g_{tt}g_{\varphi\varphi}$. By assuming a uniform specific angular momentum distribution, the term containing the specific angular momentum vanishes,
\begin{align}
    \partial_\mu(\ln|u_t|) + \frac{\partial_\mu p_{th}}{\rho h} + \frac{\partial_\mu(\mathcal{L} p_{mag})}{\mathcal{L}\rho h} = 0.
    \label{eq:potential_magnetized}
\end{align}
In order to integrate eq.~(\ref{eq:potential_magnetized}) we assume a polytropic equation of state, where the pressure is a function of the density. The thermodynamic and magnetic pressure can then be written as $p_{th} = K \rho^\Gamma$ and $p_{mag} = K_m \mathcal{L}^{q-1}(\rho h)^q$, with $K, \ K_m$ as the polytropic constants and $\Gamma, \ q$ as the polytropic exponents. The polytropic exponents will here be chosen as for relativistic degenerate matter, $\Gamma = q = \frac{4}{3}$. Using the substitutions $\tilde{p}_{mag} = \mathcal{L} p_{mag}$ and $\Tilde{\omega} = \mathcal{L} \rho h$ eq.~(\ref{eq:potential_magnetized}) can be written in integral form as
\begin{align}
     \ln|u_t| + \int^{p_{th}}_{(p_{th})_{in}} \frac{1}{\rho h} \mathrm{d}p_{th}' + \int^{\widetilde{p}_{mag}}_{(\widetilde{p}_{mag})_{in}} \frac{1}{\widetilde{\omega}} \mathrm{d} \tilde{p}_{mag}' = C. \label{eq:magnet_integral}
\end{align}
The subscript \textit{in} is referring to the inner edge of the disk.
From the equation above we identify the leading term as the effective potential of the disk, $\mathcal{W} = \ln|u_t|$. This effective potential is the combination of the gravitational and centrifugal potential. The integration constant $C$ is determined by the boundary conditions at the inner edge of the disk. Since $(p_{th})_{in} = (\tilde{p}_{mag})_{in} = 0$, the integration constant is given by the effective potential at the inner edge of the disk, $C = \ln|(u_t)|_{in} = W_{in}$. Integration of eq.~(\ref{eq:magnet_integral}) yields
\begin{align}
    \mathcal{W} - \mathcal{W}_{in} + \ln(h) + \frac{q}{q-1} K_m (\mathcal{L}\rho h)^{q-1} = 0.
    \label{eq:magnet_integrated}
\end{align}
Solving this equation for the rest-mass density is only numerically possible since it is transcendental. In order to fix the gauge, the rest-mass density will be normalized at the center of the disk to $\rho_c = 1$. Rewriting eq.~(\ref{eq:magnet_integrated}) with respect to the disk center and expressing the specific enthalpy in terms of the rest-mass density is leading to
\begin{align}
        \mathcal{W}_c - \mathcal{W}_{in} + \ln\left(1 + \frac{\Gamma K}{\Gamma - 1} \rho_c^{\Gamma-1} \right)
        + \frac{q}{q-1} \frac{K \rho_c^\Gamma}{\beta_{mc} \left(\rho_c + \frac{K \Gamma \rho_c^\Gamma}{\Gamma-1}\right)} = 0,
        \label{eq:magnet_integrated_c}
\end{align}
where we introduce the magnetization parameter $\beta_{mc} = \frac{p_{{th}_c}}{p_{{mag}_c}}$ as the ratio between thermodynamic and magnetic pressure at the center of the disk. A high magnetization parameter $(\beta_{mc} \gg 1)$ corresponds to an essentially non-magnetized disk, whereas a low magnetization parameter ($\beta_{mc} \ll 1$) corresponds to a strongly magnetized disk. A mildly magnetized disk is represented by $\beta_{mc} = 1$, where the thermodynamic pressure equals the magnetic pressure. In our computations we will use $\beta_{mc} = 10^5$, $\beta_{mc} = 1$ and $\beta_{mc} = 10^{-5}$ for representing non-magnetized, mildly magnetized and highly magnetized disks. Besides the magnetization parameter, which we will use as our variable parameter, the only unknown quantity is the polytropic constant $K$. After computation of $K$ from eq.~(\ref{eq:magnet_integrated_c}), the eq.~(\ref{eq:magnet_integrated}) can be solved for the rest-mass density. As a numerical solving algorithm we use the bisection method, with an absolute convergence error of $\epsilon < 10^{-15}$ between the last and second to last iteration step.

\section{Disk Comparison}

\begin{figure}[H]
    \centering
    \includegraphics[width=0.8\linewidth]{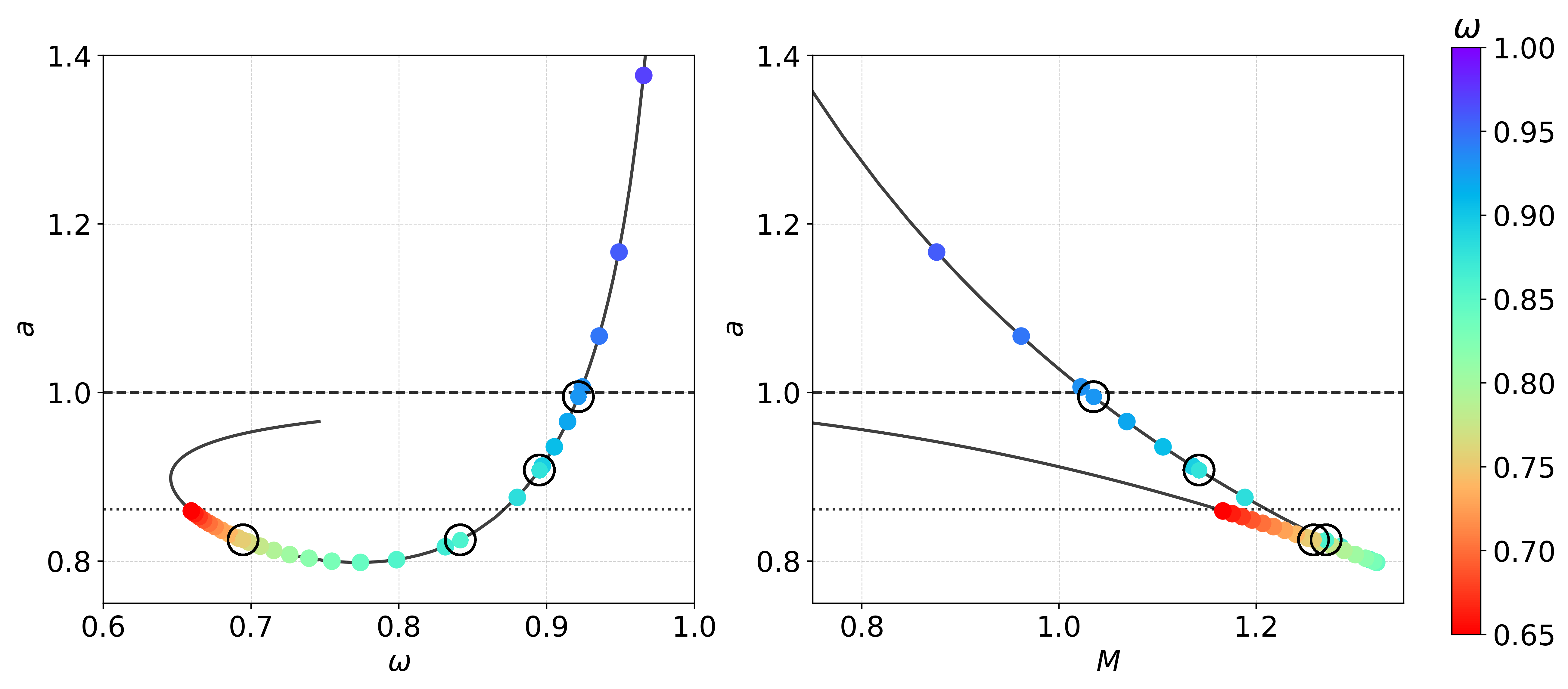}
    \caption{The left plot shows the BS spin parameter $a$ versus $\omega$. The dashed black line marks the Kerr limit $a = 1$, the dotted black line marks the value for $a$ below which the spin parameter is not monotonic for solutions without ergo-regions. The black circles highlight the chosen exemplary solutions which were  examined more closely. The right plot shows $a$ versus the BS mass $M$.}
    \label{fig:BS_spin_parameter}
\end{figure}

In order to conduct a qualitative comparison between the BS and Kerr disks, we aim to isolate the effects of magnetic fields on the disks as much as possible. Therefore we fix the other solution and disk parameters and vary the magnetization of the disk.
We employ the dimensionless spin parameter $a \equiv \frac{J}{M^2}$ for the classification of different BS and BH solutions, which are then compared to each other. The BS solution range for the spin parameter is depicted in Fig.~\ref{fig:BS_spin_parameter}. Its lower bound is $\min(a) = 0.798$, corresponding to the $\omega = 0.774$ solution, and it diverges in the limit $\omega=1$.
Since $J$ and $M$ are non-monotonic with respect to $\omega$, neither is $a$. This leads to a twofold degeneracy of the considered solutions in the interval $a \in [0.798,0.861]$. We will only focus on BS solutions in the interval $a \in [0.798,1)$, since for $a > 1$ naked singularities occur for the Kerr solutions. With the aim to examine qualitatively the various solutions classified by the spin parameter, we have selected three values of $a$, one of which represents the upper solution range, one the middle range and one the lower range, namely $a = 0.995$, $a = 0.908$ and $a = 0.825$. The spin parameter $a = 0.825$ has two corresponding BS solutions, which were therefore both examined for the sake of completeness.

\begin{figure}[H]
    \centering
    \includegraphics[width=0.8\linewidth]{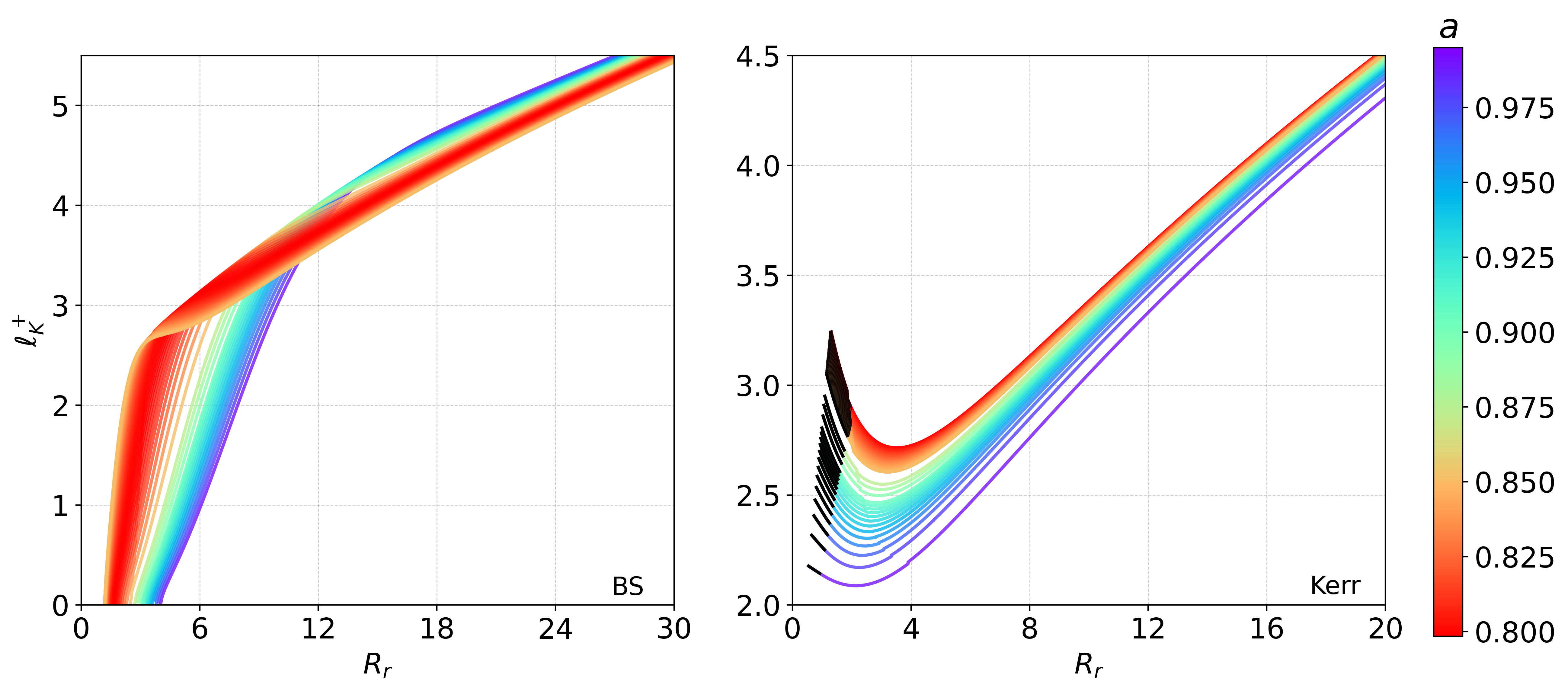}
    \caption{Prograde Keplerian specific angular momentum distribution in the equatorial plane for the various BS solutions (on the left) and Kerr black hole solutions (on the right) classified by their spin parameter $a$. The black curve sections in the Kerr plot correspond to unbound orbits.}
    \label{fig:BS_Kerr_ell+}
\end{figure}

Since we are comparing different spacetime geometries to each other, it is suitable to choose coordinates with a similar physical interpretation. Therefore we will use the proper radial distance $R_r$ as the radial coordinate for all following illustrations in the equatorial plane. $R_r$ is given by
\begin{align}
    R_r^{BS} = \int_0^r \sqrt{g_{rr}\left(r',\theta=\frac{\pi}{2} \right)} dr' \ \ ; \ \ R_r^{Kerr} = \int_{r_{H}}^{r} \sqrt{g_{rr} \left(r',\theta = \frac{\pi}{2} \right)} dr',
\end{align}
where $g_{rr}$ is the corresponding radial metric component and $r_H$ is the Kerr radial coordinate of the outer event horizon in the equatorial plane. It should be noted that $R_r$ is giving the proper radial distance to the coordinate origin for the BSs, whereas for the BHs $R_r$ is giving the proper radial distance to the outer event horizon. The origin in the following plots corresponds therefore to the horizon in the Kerr plots. For vertical and two-dimensional plots we will use pseudo-cylindrical coordinates with the proper radial distance $R_r(\theta)$ as the radial coordinate, which is then for a fixed radial value $\theta$-dependent,
\begin{align}
    R_r^{BS}(\theta) = \int_0^r \sqrt{g_{rr}\left(r',\theta \right)} dr' \ \ ; \ \ R_r^{Kerr}(\theta) = \int_{r_{H}}^{r} \sqrt{g_{rr} \left(r',\theta \right)} dr'.
\end{align}
Since all solutions possess axial symmetry, the vertical and two-dimensional plots can be done in the $x$-$z$-plane, where $x$ and $z$ are then defined as
\begin{align}
    x(r,\theta) = R_r(\theta) \sin \theta \ \ ; \ \ z(r,\theta) = R_r(\theta) \cos \theta.
\end{align}
A general analysis of the possible types of disk solutions can be done by examining the Keplerian specific angular momentum distribution, as it determines the existence of a cusp (and possibly the number of disk centers). Fig.~\ref{fig:BS_Kerr_ell+} presents the equatorial Keplerian specific angular momentum distribution for the different BS and Kerr solutions, classified by their spin parameter.

Since $\ell_K^+$ is strictly monotonic for the BSs, all disk solutions are composed of cuspless one-centered disks. Furthermore there are no marginally stable and marginally bound orbits for the BSs. This makes disk solutions with a very small angular momentum possible in contrast to BH disks. These low angular momentum solutions are also called \textit{fat tori} and differ from solutions with a higher specific angular momentum in their much further extending vertical extension.
For the Kerr solutions $\ell_K^+$ has a minimum and solutions exist only for $\ell_0 \geq \min(\ell_K^+) = \ell_K^+(r_{ms}) \equiv \ell_{ms}$, where the subscript \textit{ms} is referring to the marginally stable orbit. One-centered solutions with a cusp exist for $\ell_0$ with $\ell_{ms} < \ell_0 \leq \ell_K^+(r_{mb}) \equiv \ell_{mb}$, where the subscript \textit{mb} is referring to the marginally bound orbit. In these solutions the disk center is located in the vicinity of the event horizon. For $\ell_0 > \ell_{mb}$ only solutions without a cusp are possible. The specific angular momentum of thick disks in the vicinity of the inner edge is believed to be close to $\ell_{mb}$ \cite{FoundationsADTheory}, therefore $\ell_0$ will be set to $\ell_0 = \ell^{Kerr}_{mb}$ in all following disk calculations. This corresponds to a Kerr disk solution, where the equi-density surface through the cusp closes at spatial infinity. Regarding the BS disk calculations, the specific angular momentum value is set to the same value, in order to compare disks with the same specific angular momentum. Hence the effects of magnetic fields are isolated as much as possible, since the spin parameter and the specific angular momentum are set to the same relative value for BS and Kerr solutions. In order to represent all physical feasible solutions, the effective potential at the inner edge of the disk is set to $\mathcal{W}_{in} = 0$ in all calculations.

\begin{figure}[H]
    \centering
    \includegraphics[width=\linewidth]{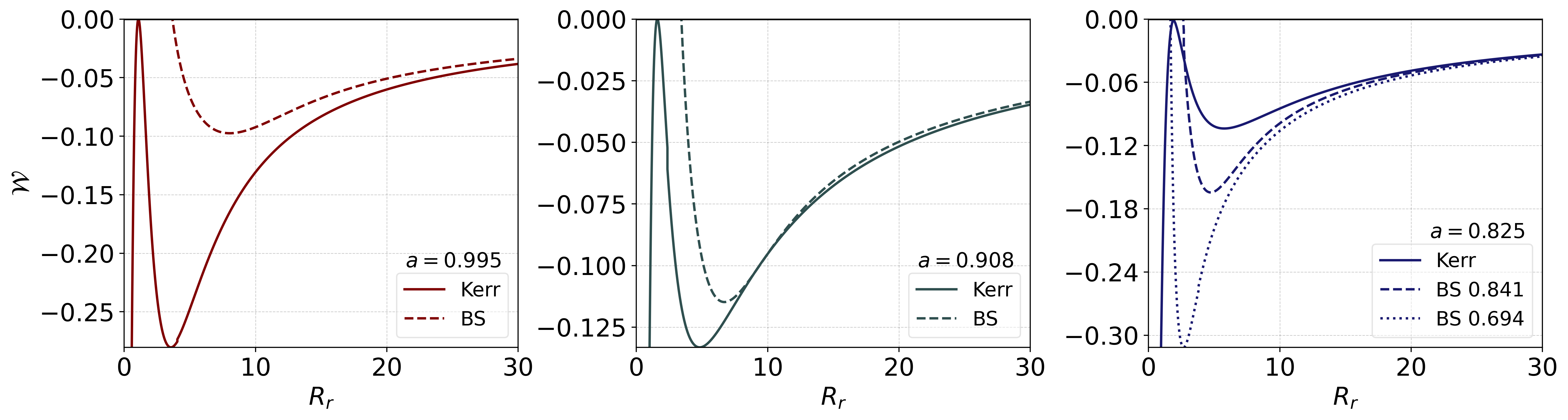}
    \caption{Effective potential $\mathcal{W}$ in the equatorial plane with the high spin parameter solutions on the left to the lower spin parameter solutions on the right. The dashed and dotted curves mark the effective potential for the BSs and the solid curves for the BHs. The minimum of the curves marks the location of the disk center, where the motion is geodesic.}
    \label{fig:eff_pot}
\end{figure}

Following the described procedure the specific angular momentum of the $a = 0.995$ solutions is set to $\ell_0 = \ell^{Kerr}_{mb} = 2.142$, for the $a = 0.908$ solutions to $\ell_0 = \ell^{Kerr}_{mb} = 2.607$ and for the $a = 0.825$ solutions to $\ell_0 = \ell^{Kerr}_{mb} = 2.836$. The corresponding effective potentials in the equatorial plane are shown in Fig.~$\ref{fig:eff_pot}$. 

For the high and mid range spin parameter solutions a test particle at the BH disk center is more energetically bound compared to the BS disks, since the minimum of the effective potential is lower for the BHs. Regarding the lower spin parameter case the opposite applies for both BSs solutions. In all three spin parameter cases $\mathcal{W}_{BS}$ is evolving similarly to $\mathcal{W}_{Kerr}$ with increasing radial distance, the biggest differences are located close to the disk center.

In the following we will first discuss the influence of magnetic fields on the disk solutions for the high spin parameter case, then we will discuss the solutions for the mid spin parameter case and finally we will discuss the small spin parameter case. For a suitable comparison between BS and BH disks we will focus primarily on the density distributions and on dimensionless quantities of the disk, which are defined and presented in Tables \ref{tab1} - \ref{tab3}.
We also examine the equatorial Bernoulli parameter of the solutions. It is a quantity particularly important for dynamical simulations, since the time evolution of the Bernoulli parameter in numerical simulations tends to be controlled by the Bernoulli parameter of the initial solution \cite{NewEquilibriumTorusSolution}. For magnetized tori it can be written as,
\begin{align}
    Be =  \left( 1 + \frac{U+p_{th}+b^2}{\rho} \right) u_t - 1,
    \label{eq:Be}
\end{align}
with $U$ being the internal energy. By using the specific internal energy $\epsilon = \frac{U}{\rho}$ and the magnetic pressure $p_{mag} = \frac{b^2}{2}$, the equation can be rewritten in terms of the magnetization parameter $\beta_m$,
\begin{align}
    Be = \left( 1 + K \rho^{\Gamma - 1} \left( \frac{\Gamma}{\Gamma - 1} + \frac{2}{\beta_m} \right) \right) u_t - 1,
\end{align}
where $\beta_m$ denotes the ratio between thermodynamic and magnetic pressure, with $\beta_m = \beta_{mc}$ at the disk center.
The Bernoulli parameter is representing the sum of the potential and kinetic energy with the gas enthalpy of the disk. It can be used as a suitable quantity to analyze the stability of disk solutions, since for steady state solutions and in the absence of viscosity it is conserved alongside streamlines. For gas at infinity $Be \geq 0$ applies, since the gravitational potential is approaching zero and the kinetic energy as well as the gas enthalpy are always positive. As a consequence any stream of disk particles with $Be \geq 0$ could potentially reach infinity and is therefore classified as unbound. In contrast to that, a flow with $Be < 0$ is called bound. Physically feasible equilibrium torus solutions are bound solutions and should therefore have a Bernoulli parameter with $Be < 0$, since unbound flows are in general more likely to cause outflows, which may affect the stability of the disk.

\newpage
\subsection{High spin parameter $a = 0.995$ ($\omega_{BS} = 0.921, \ \ell_0 = 2.142$)}
\begin{figure}[H]
\centering
\includegraphics[width=\linewidth]{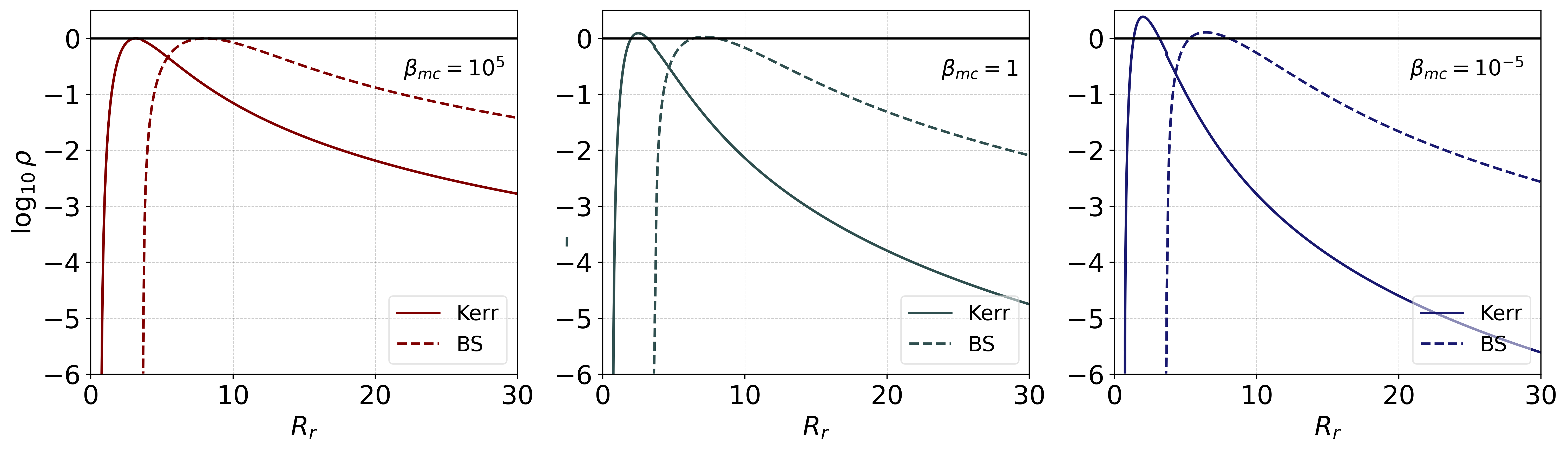}
\caption{Equatorial density distribution of the Kerr and BS disks plotted versus the proper radial coordinate with the non-magnetized, mildly magnetized and highly magnetized disks presented from left to right. The constant specific angular momentum of the disks is set to $\ell_0 = \ell^{Kerr}_{mb} = 2.142$.}
\label{fig:a995_density_eq}
\end{figure}

\begin{figure}[H]
\centering
\includegraphics[width=\linewidth]{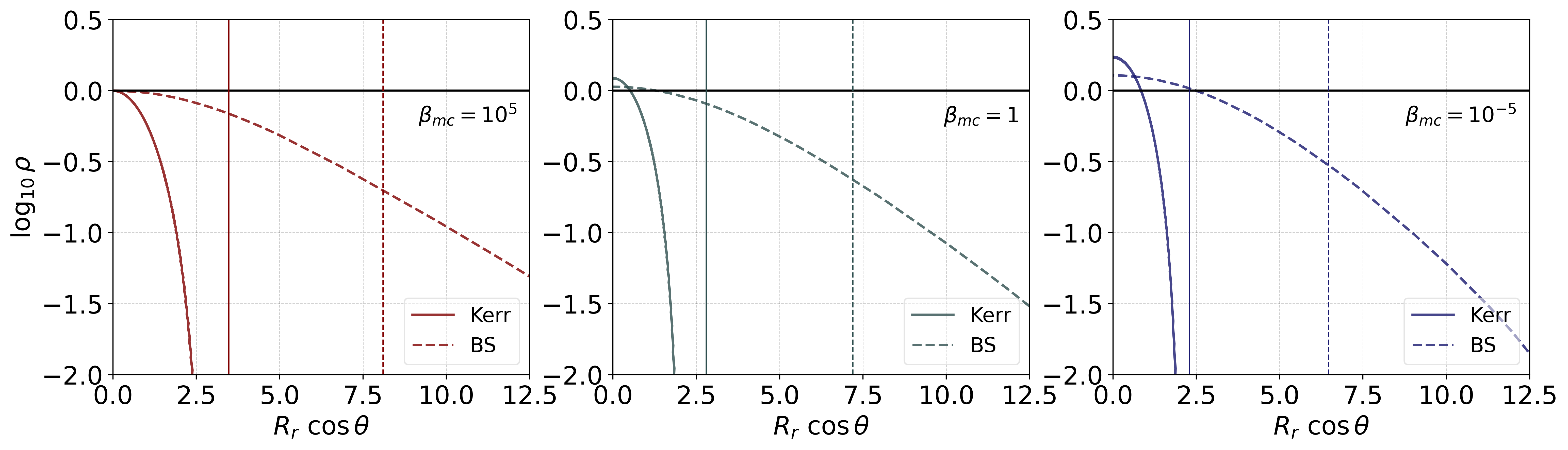}
\caption{Vertical density distribution at the density maximum of the Kerr and BS disks, with the non-magnetized, mildly magnetized and highly magnetized disks presented from left to right. The vertical lines mark the projection of the equatorial proper radial distance of the density maximum.}
\label{fig:a995_density_vertical}
\end{figure}

\begin{figure}[H]
\centering
\begin{subfigure}{.3283\textwidth}
  \centering
  \includegraphics[width=\linewidth]{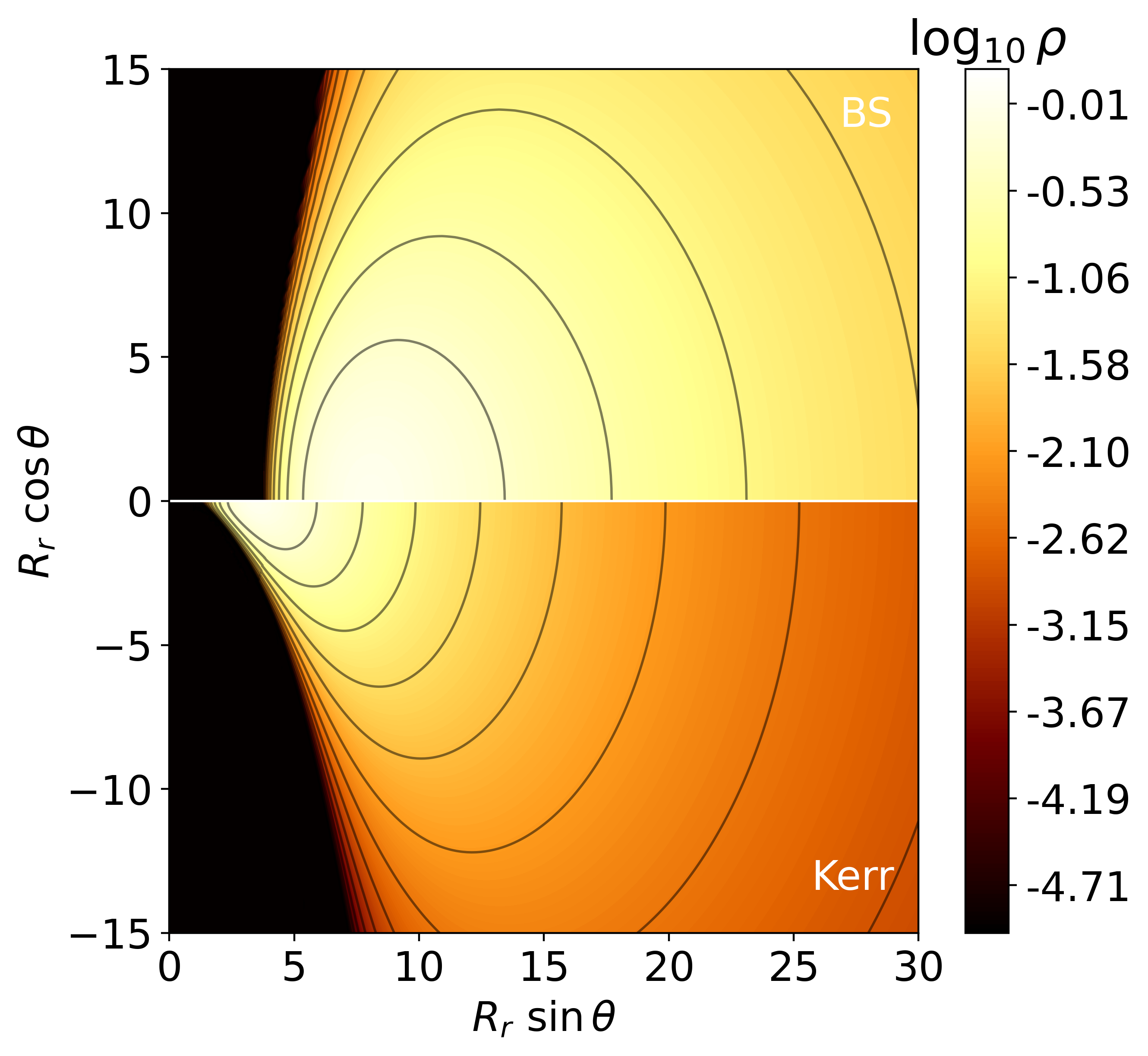}
  \caption{$\beta_{mc} = 10^5$}
\end{subfigure}
\begin{subfigure}{.3283\textwidth}
  \centering
  \includegraphics[width=\linewidth]{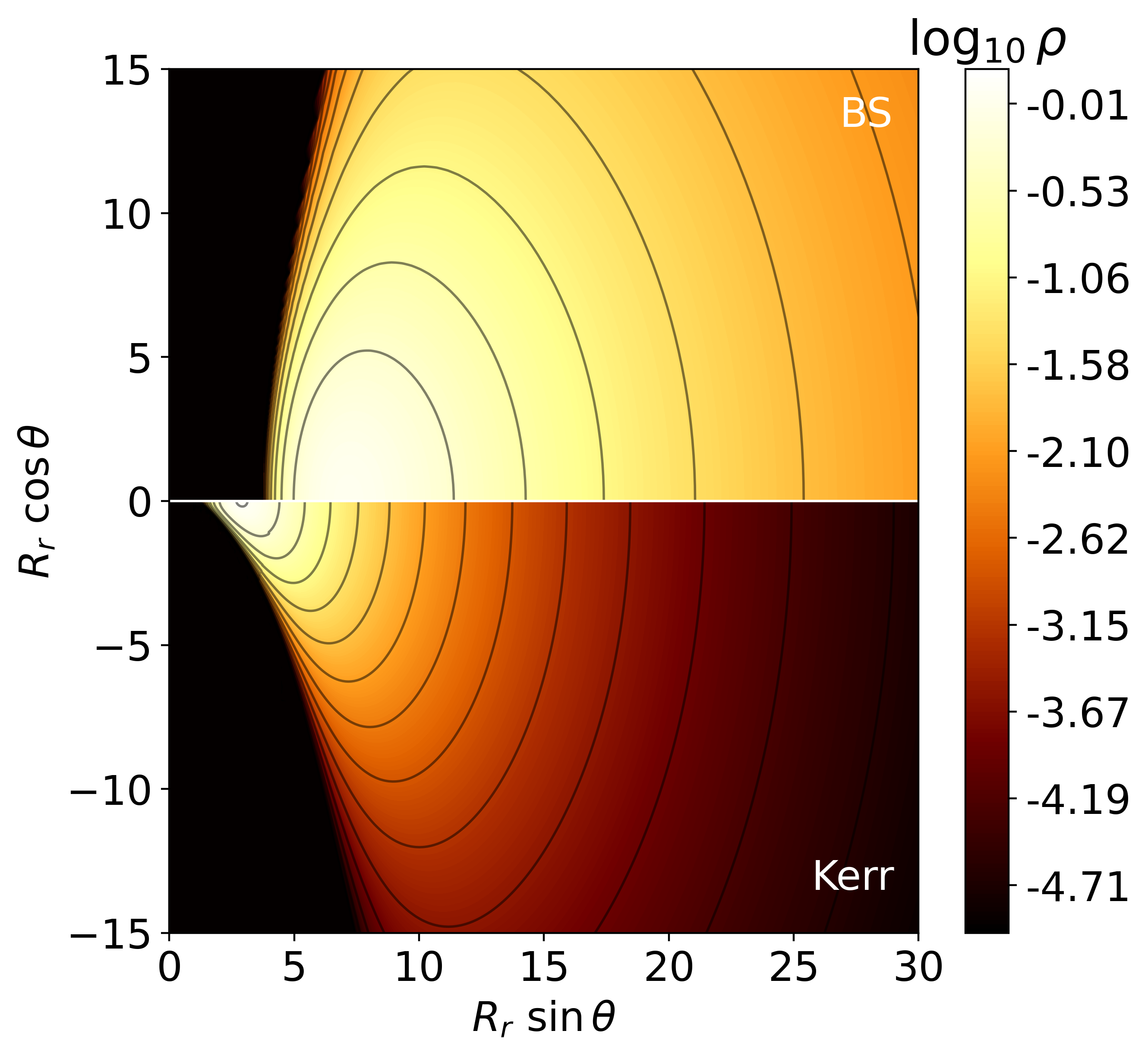}
  \caption{$\beta_{mc} = 1$}
\end{subfigure}
\begin{subfigure}{.3283\textwidth}
  \centering
  \includegraphics[width=\linewidth]{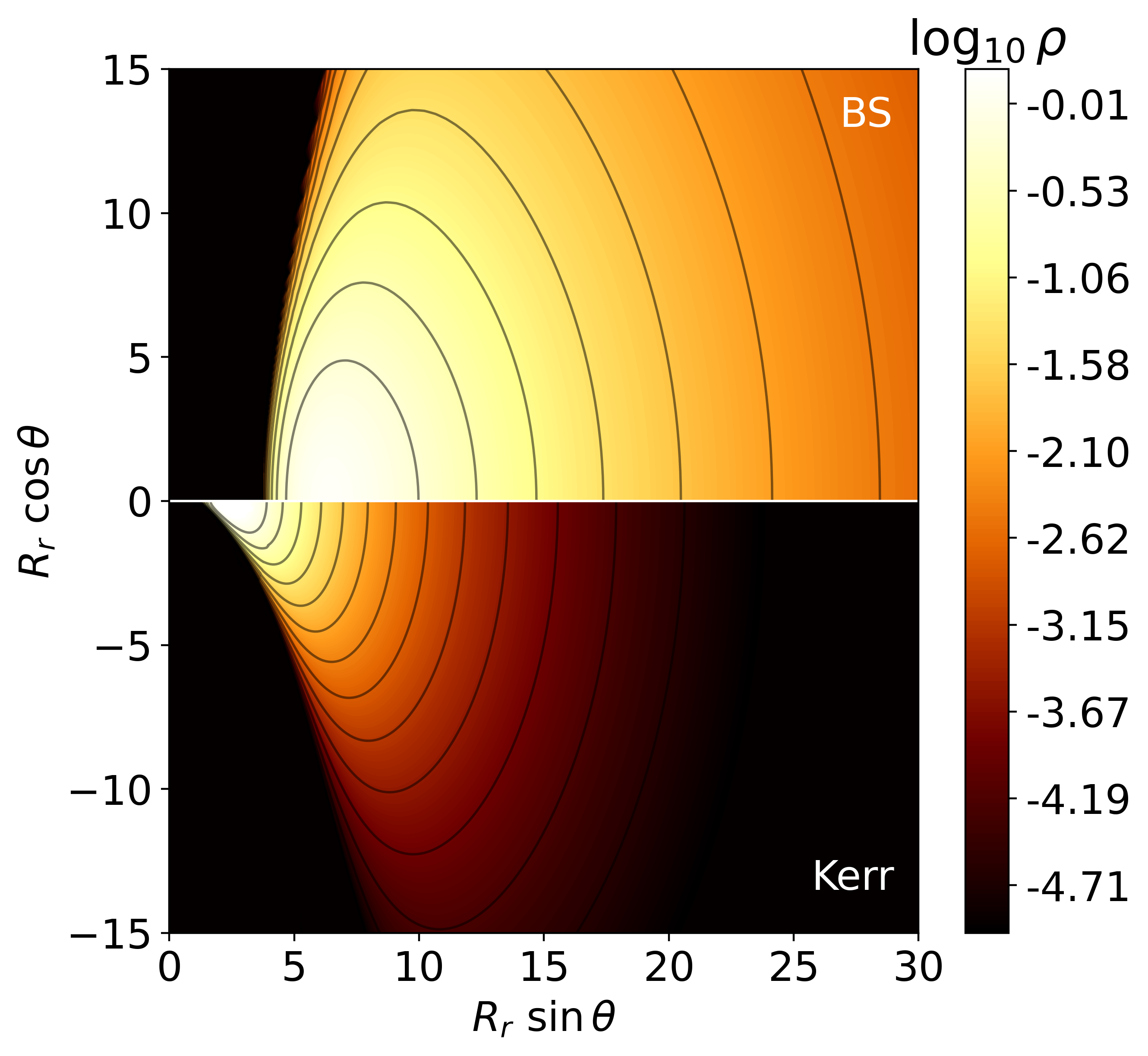}
  \caption{$\beta_{mc} = 10^{-5}$}
\end{subfigure}
\caption{Meridional cross section plots of the density distribution in pseudo-cylindrical coordinates. The upper half shows the BS torus, while the lower half shows the Kerr torus. The upper bound of the density scale is set to the maximum density between the BS and Kerr solution for the highest magnetization, $\log_{10}\rho_{\max} = 0.385$. The lower bound of the density scale is set to $\log_{10}\rho = -5$ in all plots. Black curves mark equi-density surfaces.}
\label{fig:Torus_a995}
\end{figure}

For the high spin parameter solutions the equatorial and vertical rest-mass density distributions are shown in Figs.~\ref{fig:a995_density_eq} and \ref{fig:a995_density_vertical}, and some of their characteristic properties are evaluated in Table \ref{tab1} for different magnetic fields. The density of the BS disk decreases less steeply compared to the BH disk, which is also reflected by the compactness parameter $R_{out} - R_{\rho_{max}}$ in Tab.~\ref{tab1}. For strongly magnetized disks this difference in the density distributions increases. 
The density maximum of the BS disk is located further from the origin and also further from the inner edge of the disk compared to the BH disk. The proper radial distance of the density maximum in the equatorial plane $R_{\rho_{max}}$ is more than two times higher for the BS than for the Kerr black hole in the non-magnetized case. With increasing magnetization the density maximum shifts towards the inner edge in both cases, with this shift being greater for the BH disks. The strongly magnetized BS disks are less compactified compared to the strongly magnetized BH disks, since the increase of the maximum density is greater for the Kerr disks and their vertical and radial extensions are far less outreaching. The vertical density distribution appears in both cases to be similarly affected by magnetization as the equatorial density distribution, with the greater impact on the Kerr solutions. Thus, the vertical thickness of the BH disk is far smaller and most of the mass is tightly distributed around the center in the equatorial plane, while the BS disk has a greater vertical thickness and in general is less effected by strong magnetic fields. This is also showcased in Fig.~\ref{fig:Torus_a995}, which illustrates the meridional cross-section of the density distributions. The radial extension of the disk can be quantified by evaluating the difference in the proper radial distance between the location of the density maximum in the equatorial plane $R_{\rho_{max}}$ and the equatorial cross-section of the iso-density contour which represents 10 $\%$ of the maximal density, which we denote by $R_{out}$. We see that in the non-magnetized case the radial extension of the disk is more than 2 times higher for the BS than for the Kerr black hole and the deviation increases with magnetization. As a measure of the fluid dynamics we evaluate the ratio of the angular velocity of the accreting fluid at the density maximum and the angular velocity at the inner edge of the disk. We see that the angular velocity ratio for the Kerr black hole is with 64 $\%$ higher than for the BS in the non-magnetized disks. Increasing the magnetization the angular velocity increases for the Kerr black hole while staying approximately constant for the BS. Thus, highly magnetized disks around the Kerr black hole reach with 75 $\%$ higher rotation rate than those around the BS. Since a high rotation rate is associated with improved stability against convection \cite{NewEquilibriumTorusSolution,Penna:2010}, the thick disks in the Kerr spacetime are considerably more stable compared to BSs.

\begin{figure}[H]
\centering
\includegraphics[width=\linewidth]{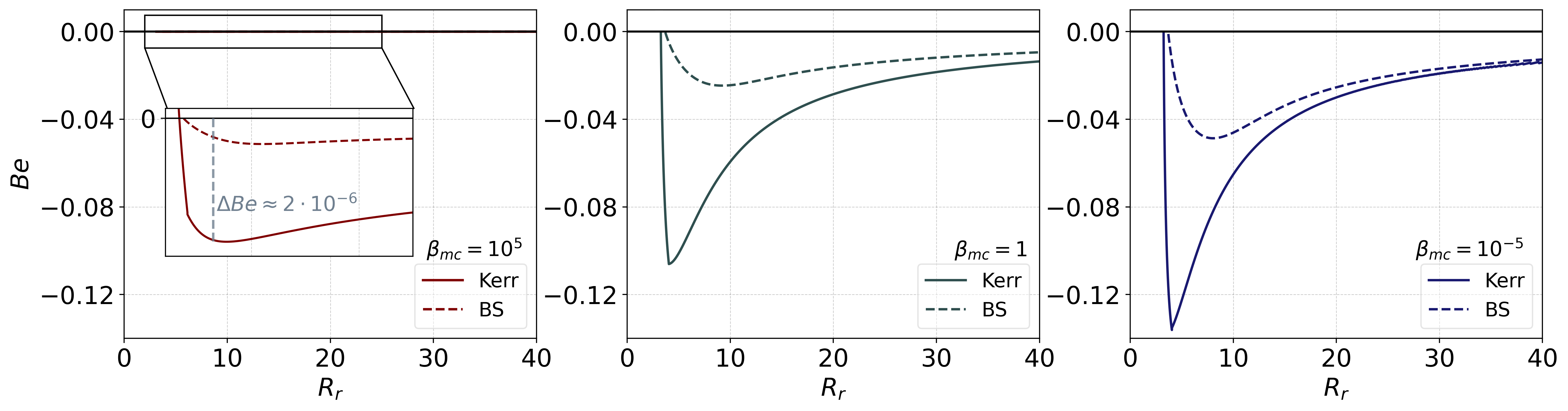}
\caption{Bernoulli parameter of the BS and Kerr disks in the equatorial plane for different degrees of magnetization, with the non-magnetized to the highly magnetized solutions from left to right. For the non-magnetized disks a zoom-in plot is inserted for a better resolution, since $Be$ is nearly constant for both disks and close to 0, with the maximum difference to 0 being on the order of $10^{-6}$.}
\label{fig:a995_Be}
\end{figure}

The Bernoulli parameter $Be$ in the equatorial plane is shown in Fig.~\ref{fig:a995_Be}. It is for the non-magnetized disks nearly constant in both cases, while for the magnetized solutions $Be$ is highly affected and develops a clear minimum, with a lower peak for the BH disk. This implies that gas and plasma flows are more energetically bound in the BH disk and are less affected by magnetic fields in the BS disk, which is in accordance to the density distributions being less affected by magnetization. Furthermore we conclude that the vertical profile and the far greater equatorial range are the main distinctive features between the BS and BH disks. \\

\begin{table}[H]
\centering
\vline
\begin{tabular}{c|c|c|c|c|c}
     \toprule
      $\beta_{mc} = 10^5$ & $\rho_{max}$ & $R_{\rho_{max}}$ & $R_{out} - R_{\rho_{max}}$ &$\frac{R_{in}}{R_{\rho_{max}}}$ & $\frac{\Omega_{\rho_{max}}}{\Omega_{in}}$ \\
     \midrule
     Kerr & 1 & 3.474 & 6.245 & 0.285 & 0.663 \\
     BS 0.921 & 1 & 8.106 & 13.883 & 0.464 & 0.240 \\
     \bottomrule
\end{tabular}
\vline
\\
\vline
\begin{tabular}{c|c|c|c|c|c|c}
     \toprule
      $\beta_{mc} = 1$ & $\rho_{max}$ & $R_{\rho_{max}}$ & $R_{out} - R_{\rho_{max}}$ & $\Delta R_{rel}$ & $\frac{R_{in}}{R_{\rho_{max}}}$ & $\frac{\Omega_{\rho_{max}}}{\Omega_{in}}$ \\
     \midrule
     Kerr & 1.236 & 2.798 & 3.446 & 0.195 & 0.354 & 0.776 \\
     BS 0.921 & 1.065 & 7.197 & 9.546 & 0.112 & 0.523  & 0.232 \\
     \bottomrule
\end{tabular}
\vline
\\
\vline
\begin{tabular}{c|c|c|c|c|c|c}
     \toprule
     $\beta_{mc} = 10^{-5}$ & $\rho_{max}$ & $R_{\rho_{max}}$ & $R_{out} - R_{\rho_{max}}$ & $\Delta R_{rel}$ & $\frac{R_{in}}{R_{\rho_{max}}}$ & $\frac{\Omega_{\rho_{max}}}{\Omega_{in}}$ \\
     \midrule
     Kerr & 2.424 & 2.288 & 2.408 & 0.341 & 0.433 & 0.856 \\
     BS 0.921 & 1.281 & 6.467 & 7.650 & 0.202 & 0.582 & 0.218 \\
     \bottomrule
\end{tabular}
\vline
\caption{Selected properties of the Kerr and BS disks in the equatorial plane, where $\rho_{max}$ is the maximum density, $R_{\rho_{max}}$ is the location of the density maximum, $R_{out} - R_{\rho_{max}}$ is the equatorial distance from the density maximum to the effective outer edge of the disk $R_{out} \coloneqq \{ R_r : \rho \left(R_r,\theta = \frac{\pi}{2}\right) = 0.1 \cdot \rho_{max}\}$, $\Delta R_{rel}$ is the relative difference in the location of the density maximum between magnetized and non-magnetized disk, $\frac{R_{in}}{R_{\rho_{max}}}$ is the ratio between inner edge and density maximum location and $\frac{\Omega_{\rho_{max}}}{\Omega_{in}}$ is the ratio between the angular velocity at the density maximum and the inner edge of the disk.}
\label{tab1}
\end{table}

\newpage
\subsection{Medium spin parameter $a = 0.908$ ($\omega_{BS} =  0.895, \ \ell_0 = 2.607$)}

\begin{figure}[H]
\centering
\includegraphics[width=\linewidth]{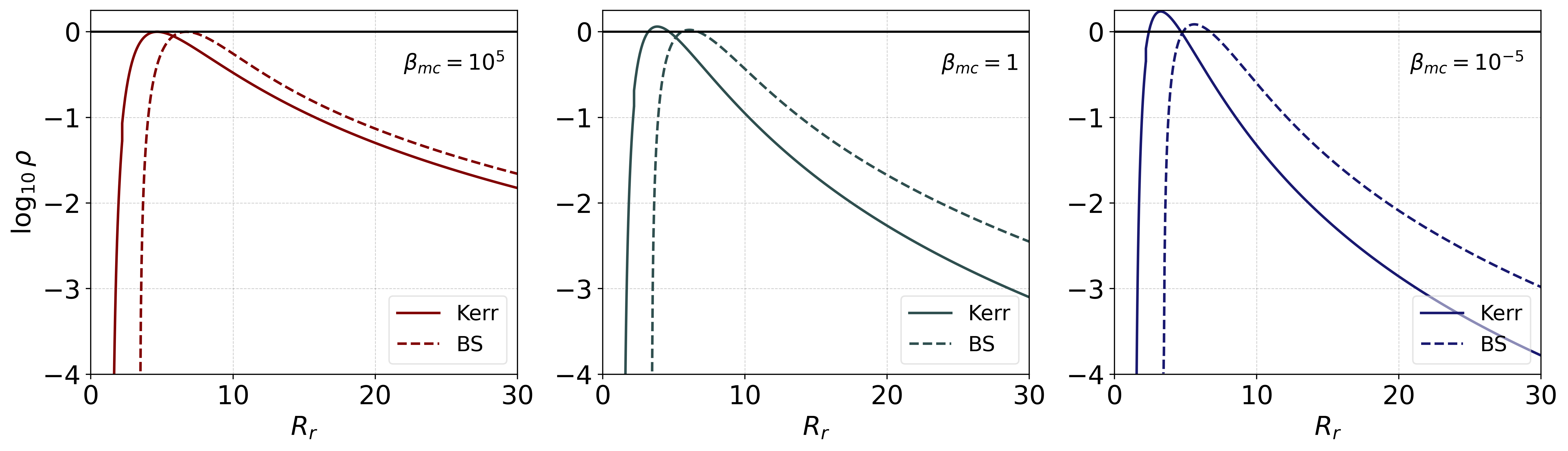}
\caption{Equatorial density distribution of the Kerr and BS disks plotted versus the proper radial coordinate with the non-magnetized, mildly magnetized and highly magnetized disks presented from left to right. The constant specific angular momentum of the disks is $\ell_0 = \ell^{Kerr}_{mb} = 2.607$}
\label{fig:a908_density_eq}
\end{figure}

\begin{figure}[H]
\centering
\includegraphics[width=\linewidth]{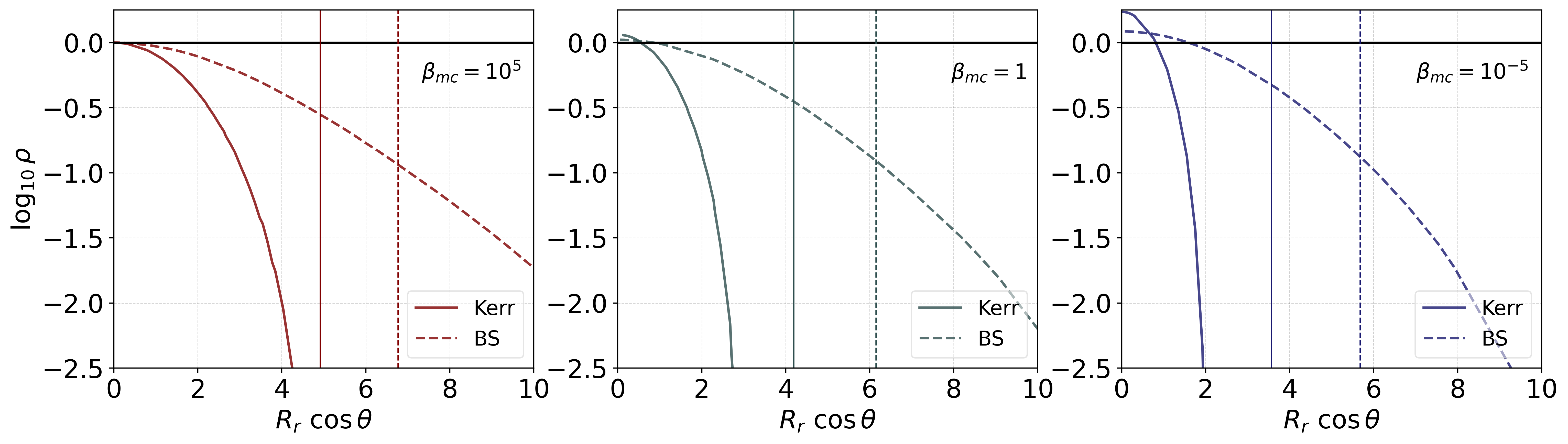}
\caption{Vertical density distribution at the density maximum of the Kerr and BS disks, with the non-magnetized, mildly magnetized and highly magnetized disks presented from left to right. The vertical lines mark the projection of the equatorial proper radial distance of the density maximum.}
\label{fig:a908_density_vertical}
\end{figure}

\begin{figure}[H]
\centering
\begin{subfigure}{.3283\textwidth}
  \centering
  \includegraphics[width=\linewidth]{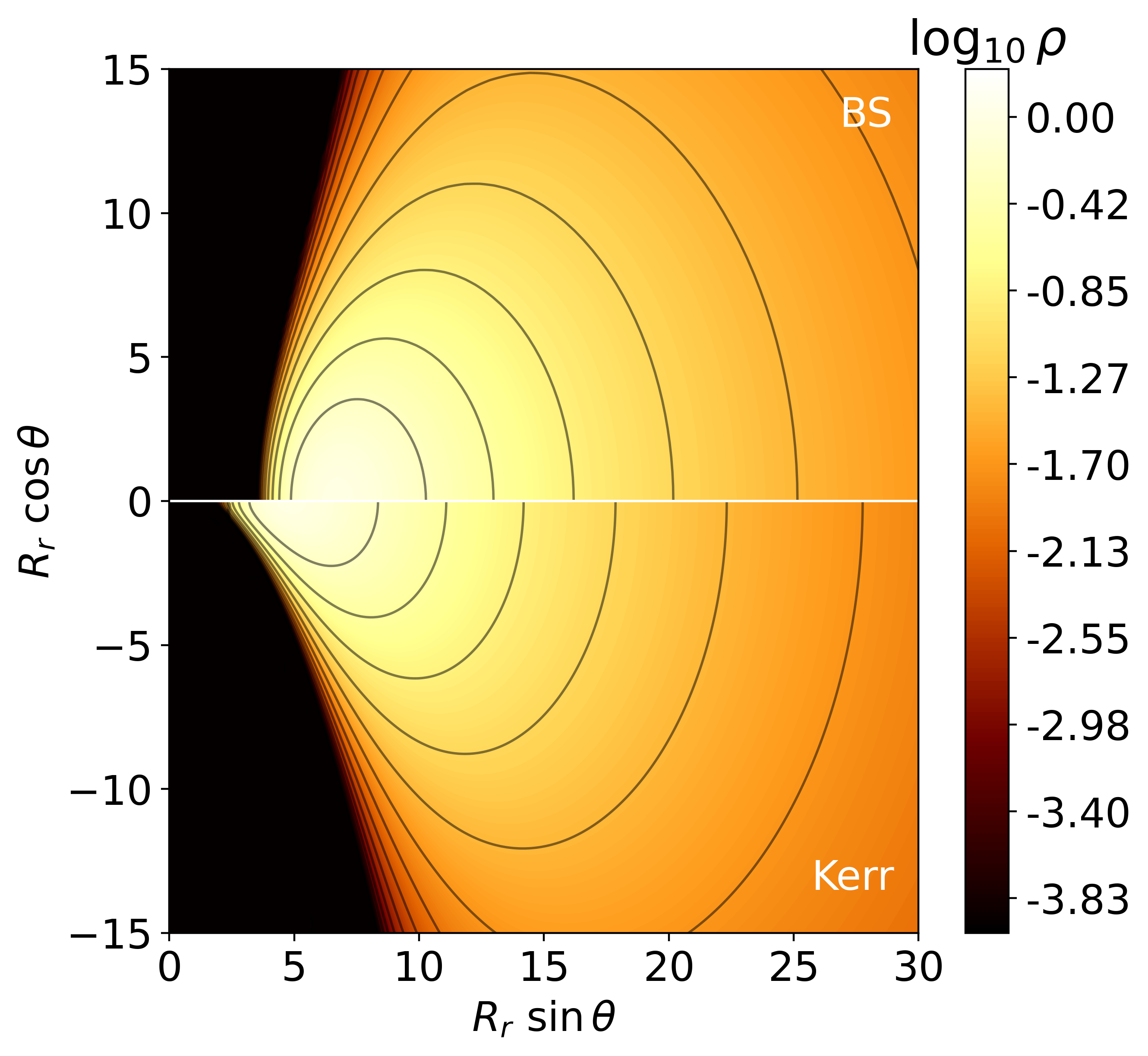}
  \caption{$\beta_{mc} = 10^5$}
\end{subfigure}
\begin{subfigure}{.3283\textwidth}
  \centering
  \includegraphics[width=\linewidth]{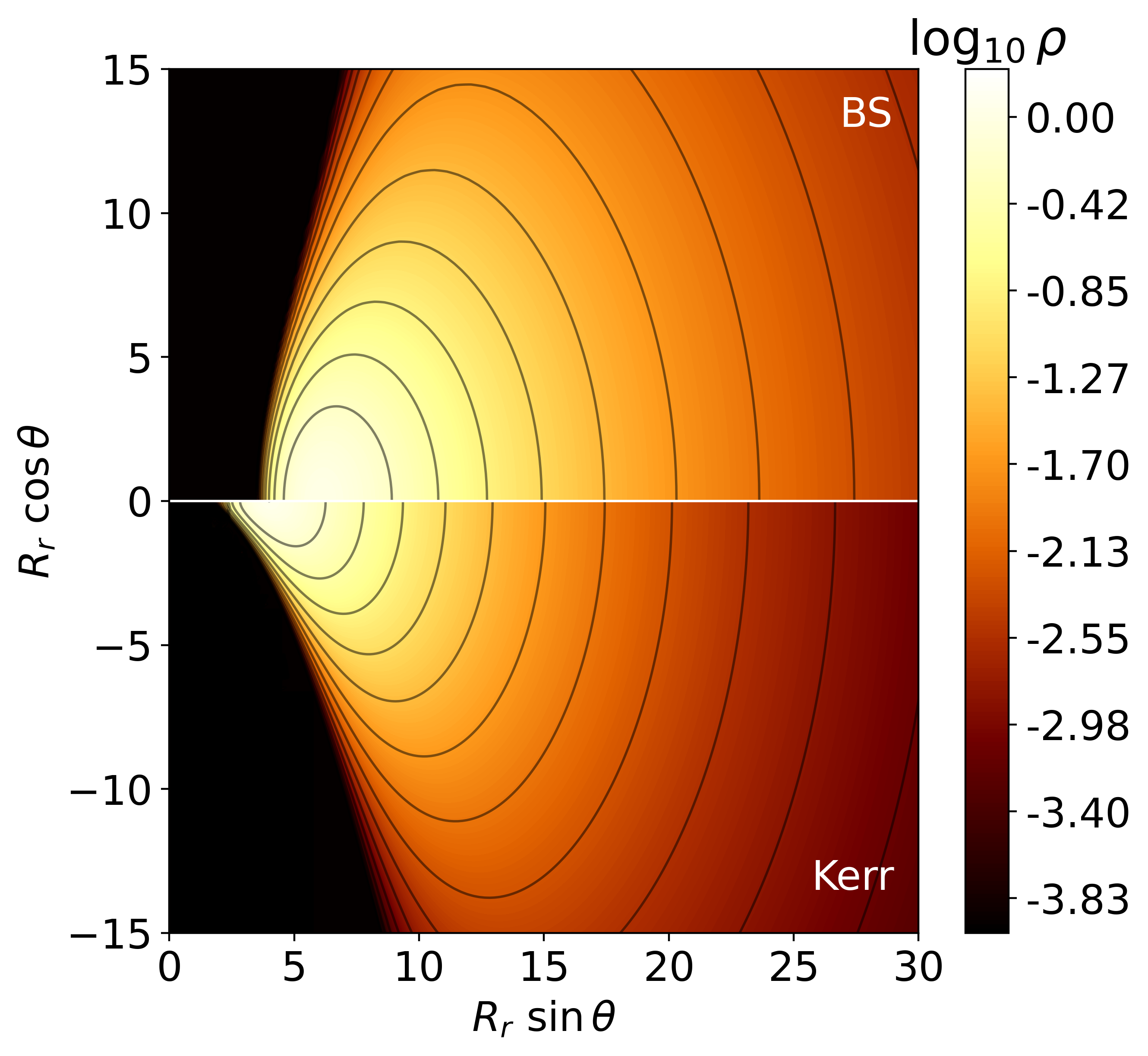}
  \caption{$\beta_{mc} = 1$}
\end{subfigure}
\begin{subfigure}{.3283\textwidth}
  \centering
  \includegraphics[width=\linewidth]{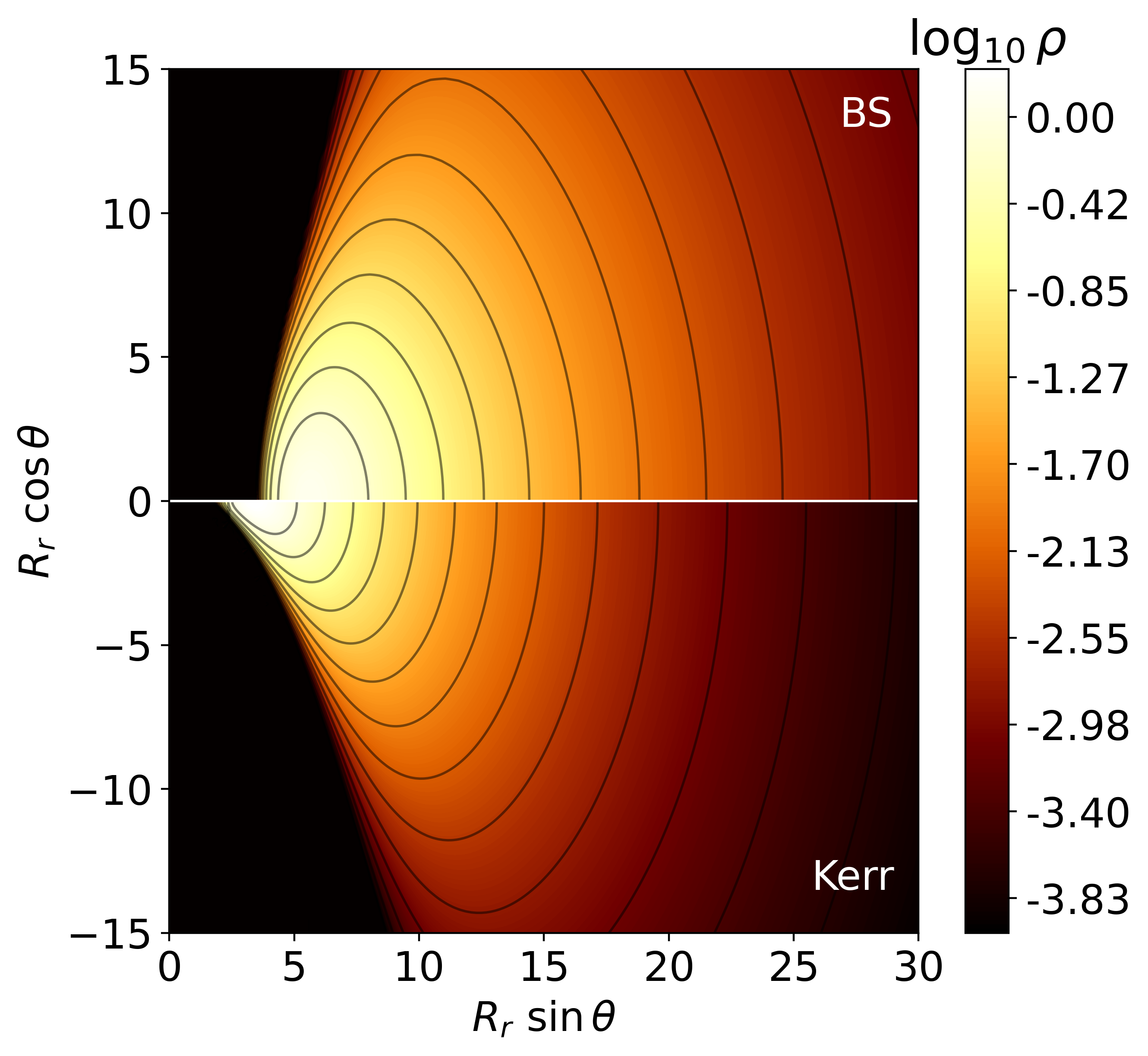}
  \caption{$\beta_{mc} = 10^{-5}$}
\end{subfigure}
\caption{Meridional cross section plots of the density distribution in pseudo-cylindrical coordinates. The upper half shows the BS torus, while the lower half shows the Kerr torus. The upper bound of the density scale is set to the maximum density between the BS and Kerr solution for the highest magnetization, $\log_{10}\rho_{\max} = 0.238$. The lower bound of the density scale is set to $\log_{10}\rho = -4$ in all plots. Black curves mark equi-density surfaces.}
\label{fig:Torus_a908}
\end{figure}

The equatorial and vertical rest-mass density distributions are depicted in Figs.~\ref{fig:a908_density_eq} and \ref{fig:a908_density_vertical}, and some characteristic quantities are given in Table \ref{tab2}. Compared to the high spin parameter solutions, the slope of the equatorial density curves are more similar, the difference in the density decrease close to the center is less than an order of magnitude for all degrees of magnetization. As for the high spin parameter solutions the density curve has a less steep slope for the BS disks for all magnetization parameters and the density maximum of the BH disks is located closer to the origin. The relative change in the location of the density maximum is greater for the BHs with increasing magnetization, and in general the BH disks get more compactified resulting in a denser center. For the non-magnetized disk the proper radial distance of the density maximum is with 38 $\%$ higher for the BS compared to the Kerr black hole. Increasing the magnetization the deviation grows reaching 65 $\%$ for highly magnetized disks. The vertical thickness of the BS disk is greater for all cases, since the vertical density curve for the BH disks has again a steeper slope. This difference increases for the highly magnetized disks, where the BS vertical density profile is largely unaffected and the BH vertical profile drops slightly sharper compared to the non-magnetized disk. These differences in the density distribution are less profound than in the high spin parameter case, as the cross-section plots in Fig.~\ref{fig:Torus_a908} illustrate. The thick disks around black holes rotate faster than those around boson stars irrespective of magnetization. In the non-magnetized case the relative angular velocity at the density maximum is with 27 $\%$ higher for BHs than for BSs. Increasing the magnetization the angular velocity for the Kerr black hole grows while for BSs it remains approximately constant. As a result, for highly magnetized disks the deviation between the two spacetimes reaches 52 $\%$.

\begin{figure}[H]
\centering
\includegraphics[width=\linewidth]{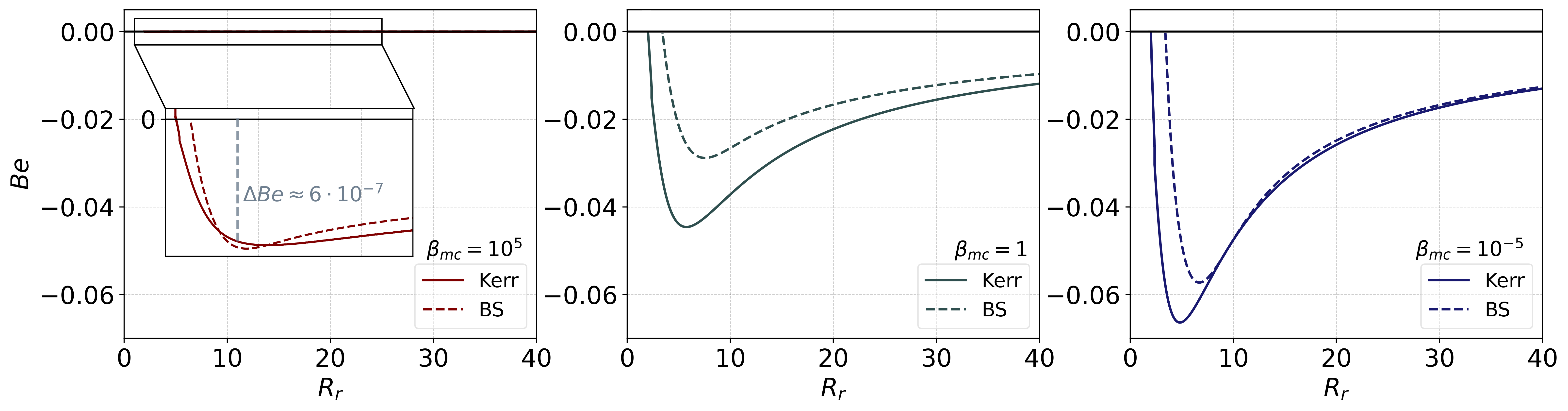}
\caption{Bernoulli parameter of the BS and Kerr disks in the equatorial plane for different degrees of magnetization, with the non-magnetized to the high magnetized solutions from left to right. For the non-magnetized disks a zoom-in plot is inserted for a better resolution, since $Be$ is nearly constant for both disks and close to 0, with the maximum difference to 0 being on the order of $10^{-7}$.}
\label{fig:a908_Be}
\end{figure}

The Bernoulli parameter in the equatorial plane (Fig.~\ref{fig:a908_Be}) behaves similarly as in the high spin parameter solutions. For the non-magnetized disks it is nearly constant and for the magnetized disks it has a clear minimum, with a lower peak for the BS disk. Since $Be$ behaves similarly for BS and Kerr, the disks are comparably energetically bound. We conclude that the BS and BH disks remain similar even under the influence of strong magnetic fields, whereby the magnetized BH disks are again characterized by a higher degree of compactification and the magnetized BS disks by their largely unaffected vertical thickness. \\

\begin{table}[H]
\centering
\vline
\begin{tabular}{c|c|c|c|c|c}
     \toprule
      $\beta_{mc} = 10^5 $& $\rho_{max}$ & $R_{\rho_{max}}$ & $R_{out} - R_{\rho_{max}}$ & $\frac{R_{in}}{R_{\rho_{max}}}$ & $\frac{\Omega_{\rho_{max}}}{\Omega_{in}}$ \\
     \midrule
     Kerr & 1 & 4.919 & 11.292 & 0.324 & 0.416 \\
     BS 0.895 & 1 & 6.769 & 11.364 & 0.513 & 0.304\\
     \bottomrule
\end{tabular}
\vline
\\
\vline
\begin{tabular}{c|c|c|c|c|c|c}
     \toprule
      $\beta_{mc} = 1$& $\rho_{max}$ & $R_{\rho_{max}}$ & $R_{out} - R_{\rho_{max}}$ & $\Delta R_{rel}$ & $\frac{R_{in}}{R_{\rho_{max}}}$ & $\frac{\Omega_{\rho_{max}}}{\Omega_{in}}$ \\
     \midrule
     Kerr & 1.150 & 4.041 & 6.197 & 0.179 & 0.394 & 0.527 \\
     BS 0.895 & 1.053 & 6.155 & 7.604 & 0.091 & 0.564  & 0.303 \\
     \bottomrule
\end{tabular}
\vline
\\
\vline
\begin{tabular}{c|c|c|c|c|c|c}
     \toprule
     $\beta_{mc} = 10^{-5}$ & $\rho_{max}$ & $R_{\rho_{max}}$ & $R_{out} - R_{\rho_{max}}$ & $\Delta R_{rel}$ & $\frac{R_{in}}{R_{\rho_{max}}}$ & $\frac{\Omega_{\rho_{max}}}{\Omega_{in}}$ \\
     \midrule
     Kerr & 1.727 & 3.481 & 4.348 & 0.292 & 0.458 & 0.616 \\
     BS 0.895 & 1.222 & 5.680 & 6.035 & 0.161 & 0.611 & 0.298 \\
     \bottomrule
\end{tabular}
\vline
\caption{Selected properties of the Kerr and BS disks in the equatorial plane, where $\rho_{max}$ is the maximum density, $R_{\rho_{max}}$ is the location of the density maximum,  $R_{out} - R_{\rho_{max}}$ is the equatorial distance from the density maximum to the effective outer edge of the disk, $\Delta R_{rel}$ is the relative difference in the location of the density maximum between magnetized and non-magnetized disk, $\frac{R_{in}}{R_{\rho_{max}}}$ is the ratio between inner edge and density maximum location and $\frac{\Omega_{\rho_{max}}}{\Omega_{in}}$ is the ratio between the angular velocity at the density maximum and the inner edge of the disk.}
\label{tab2}
\end{table}

\newpage
\subsection{Lower spin parameter $a = 0.825$ ($\omega_{BS} = \{0.694, 0.841\}, \ \ell_0 = 2.836$)}

\begin{figure}[H]
\centering
\includegraphics[width=\linewidth]{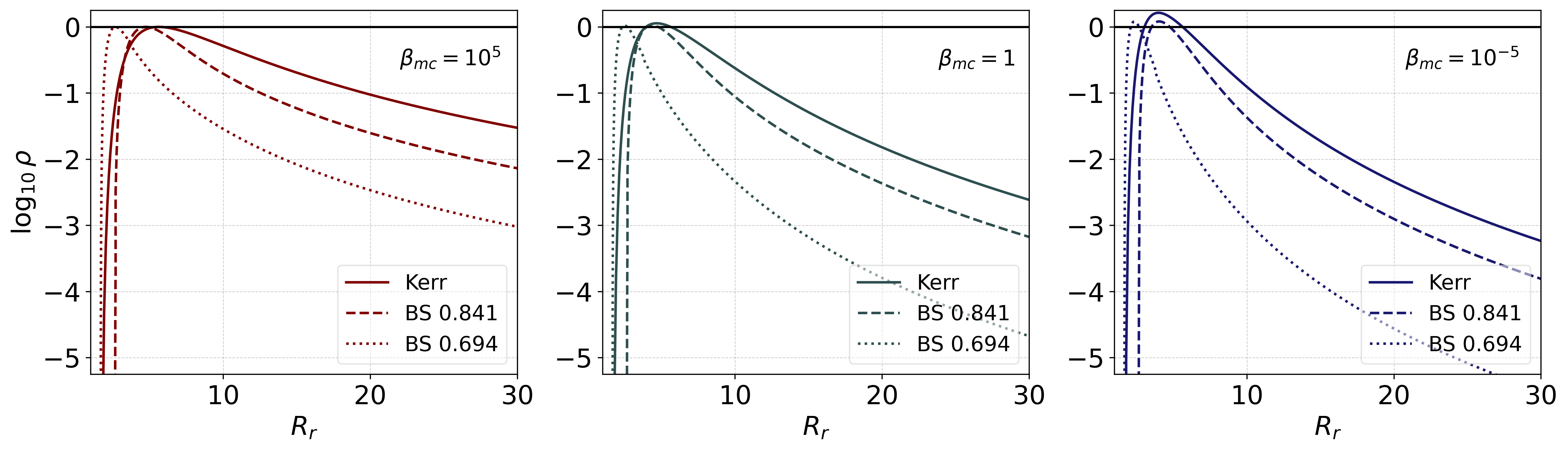}
\caption{Equatorial density distribution of the Kerr and BS disks for the non-magnetized, mildly magnetized and highly magnetized presented case from left to right. The constant specific angular momentum of the disks is $\ell_0 = \ell^{Kerr}_{mb} = 2.836$.}
\label{fig:a825_density_eq}
\end{figure}

\begin{figure}[H]
\centering
\includegraphics[width=\linewidth]{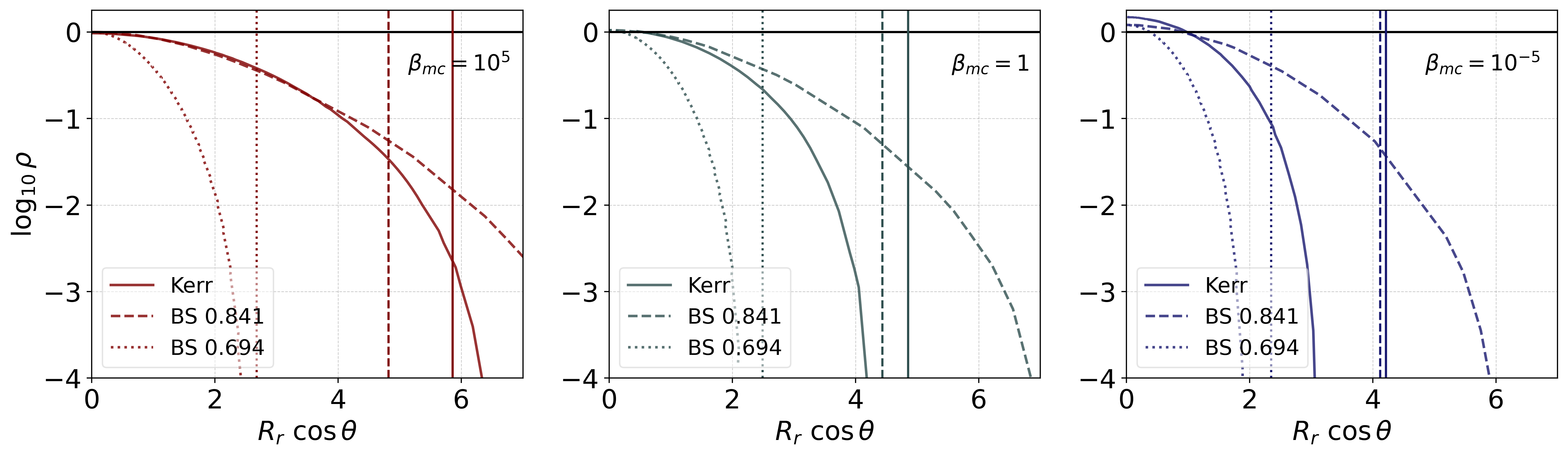}
\caption{Vertical density distribution at the density maximum of the Kerr and BS disks, with the non-magnetized, mildly magnetized and highly magnetized disks depicted from left to right. The vertical lines represent the location of the respective density maximum in the equatorial plane.}
\label{fig:a825_density_vertical}
\end{figure}

For the lower spin parameter there are two BS solutions with $\omega_{BS} = 0.694$ and $\omega_{BS} = 0.841$. The former represents a highly relativistic boson star while the higher value of the angular frequency $\omega_{BS}$ corresponds to a more mildly relativistic solution (see Fig.~\ref{fig:BS_solutions}). The equatorial and vertical rest-mass density distributions are depicted in Figs.~\ref{fig:a825_density_eq} and \ref{fig:a825_density_vertical}, while some of their properties are evaluated in Table \ref{tab3}.

\begin{figure}[H]
\centering
\begin{subfigure}{.3283\textwidth}
  \centering
  \includegraphics[width=\linewidth]{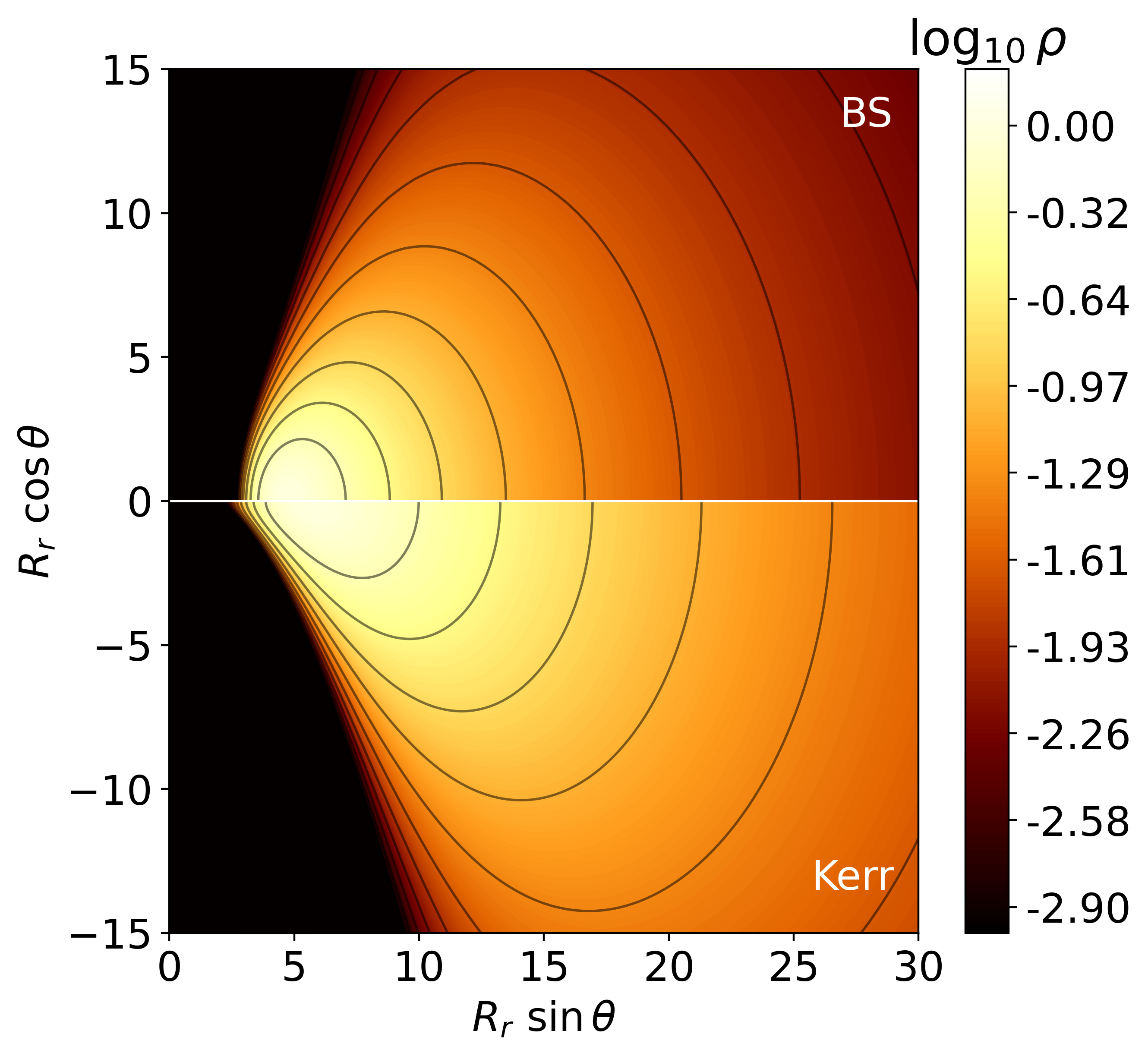}
  \caption{BS: 0.841, $\beta_{mc} = 10^5$}
\end{subfigure}
\begin{subfigure}{.3283\textwidth}
  \centering
  \includegraphics[width=\linewidth]{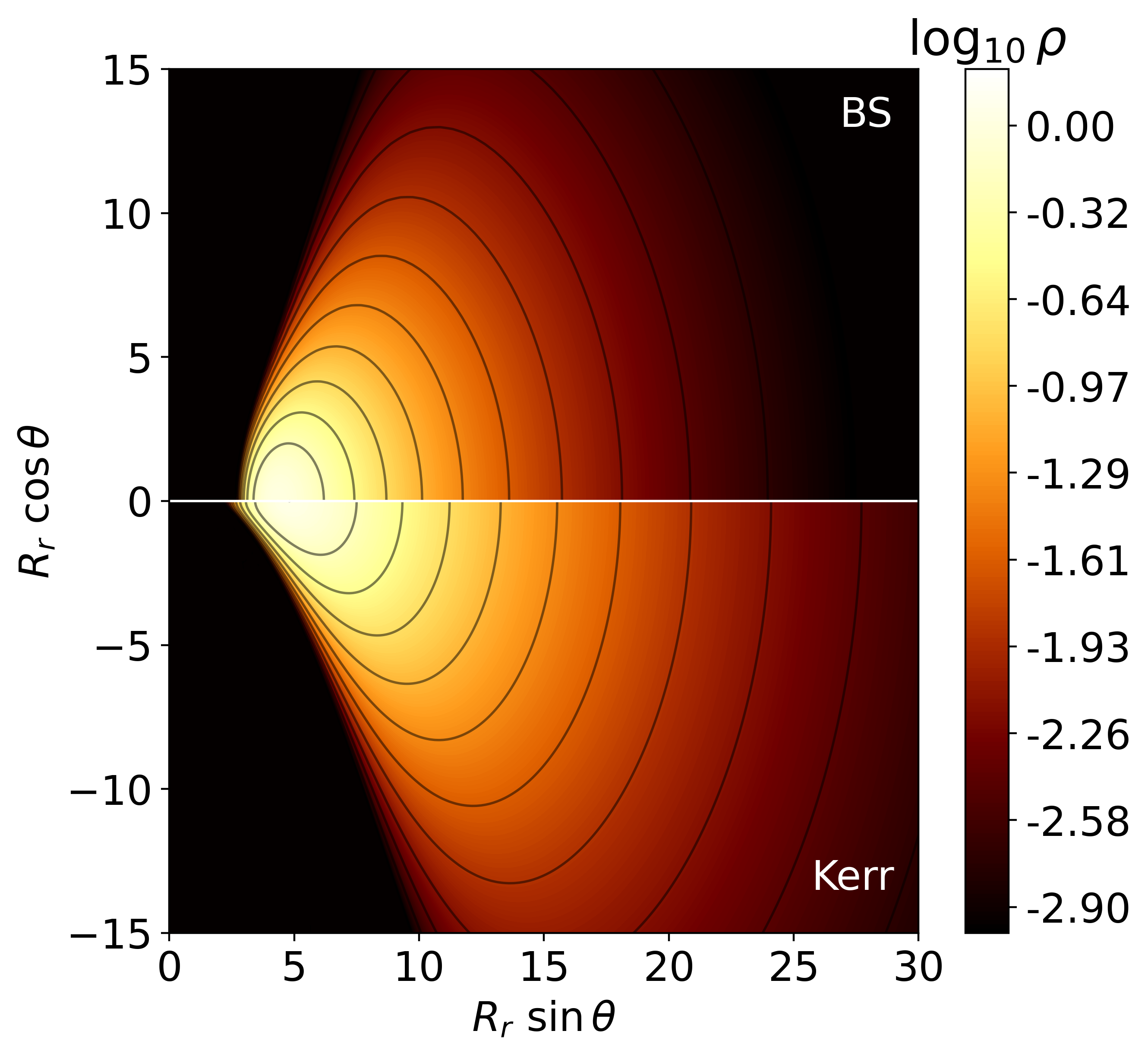}
  \caption{BS: 0.841, $\beta_{mc} = 1$}
\end{subfigure}
\begin{subfigure}{.3283\textwidth}
  \centering
  \includegraphics[width=\linewidth]{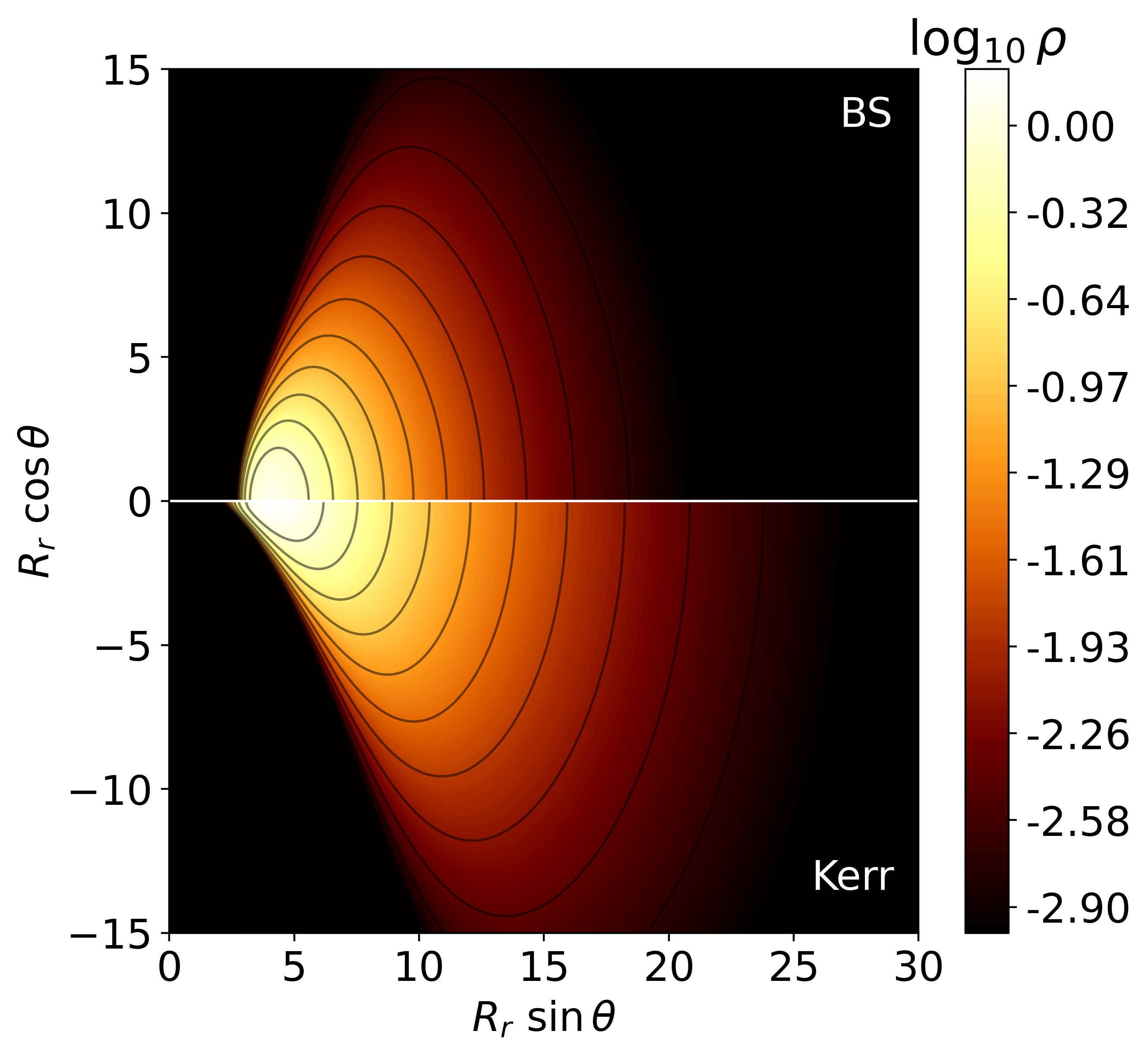}
  \caption{BS: 0.841, $\beta_{mc} = 10^{-5}$}
\end{subfigure}
\begin{subfigure}{.3283\textwidth}
  \centering
  \includegraphics[width=\linewidth]{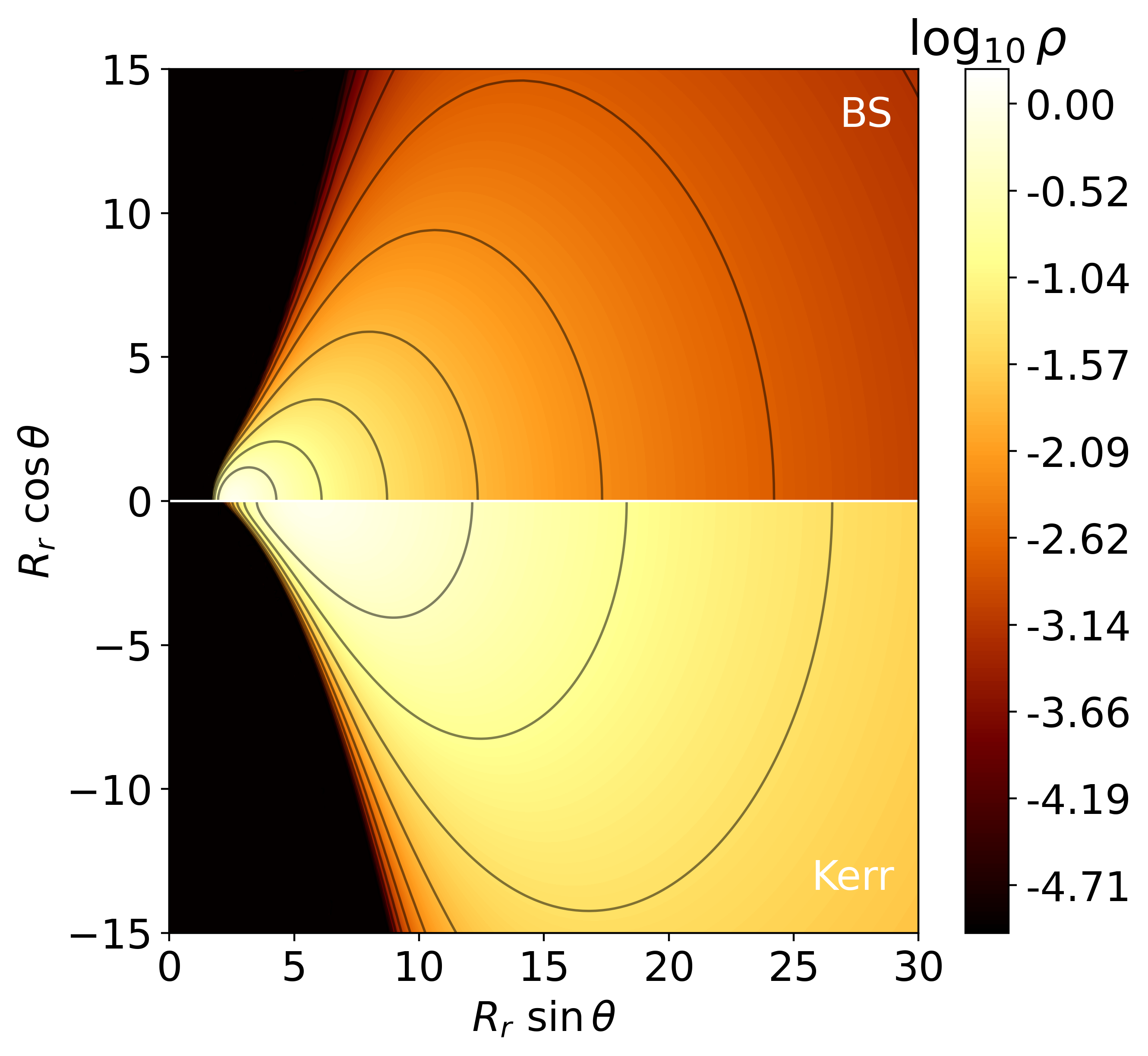}
  \caption{BS: 0.694, $\beta_{mc} = 10^5$}
\end{subfigure}
\begin{subfigure}{.3283\textwidth}
  \centering
  \includegraphics[width=\linewidth]{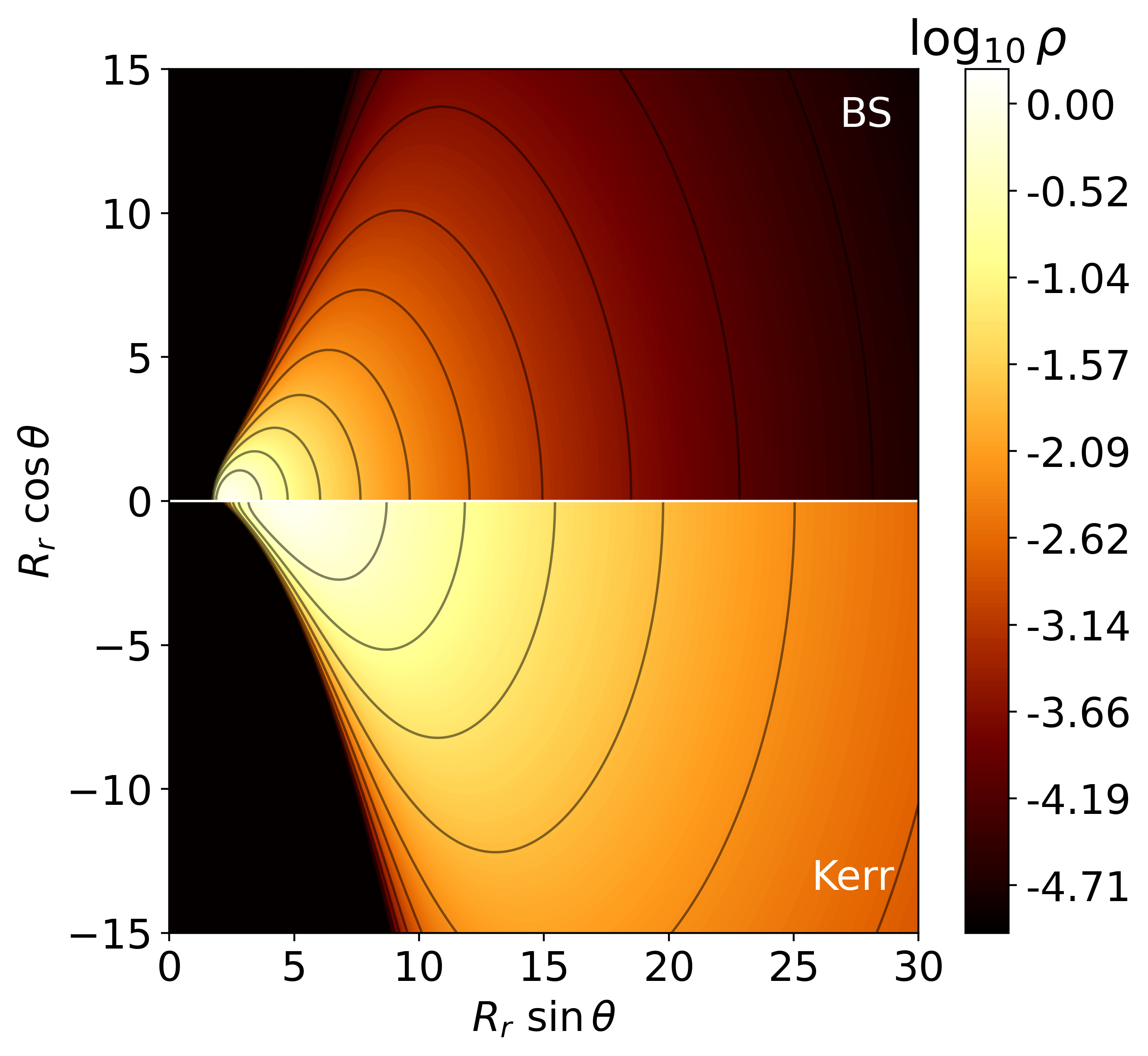}
  \caption{BS: 0.694, $\beta_{mc} = 1$}
\end{subfigure}
\begin{subfigure}{.3283\textwidth}
  \centering
  \includegraphics[width=\linewidth]{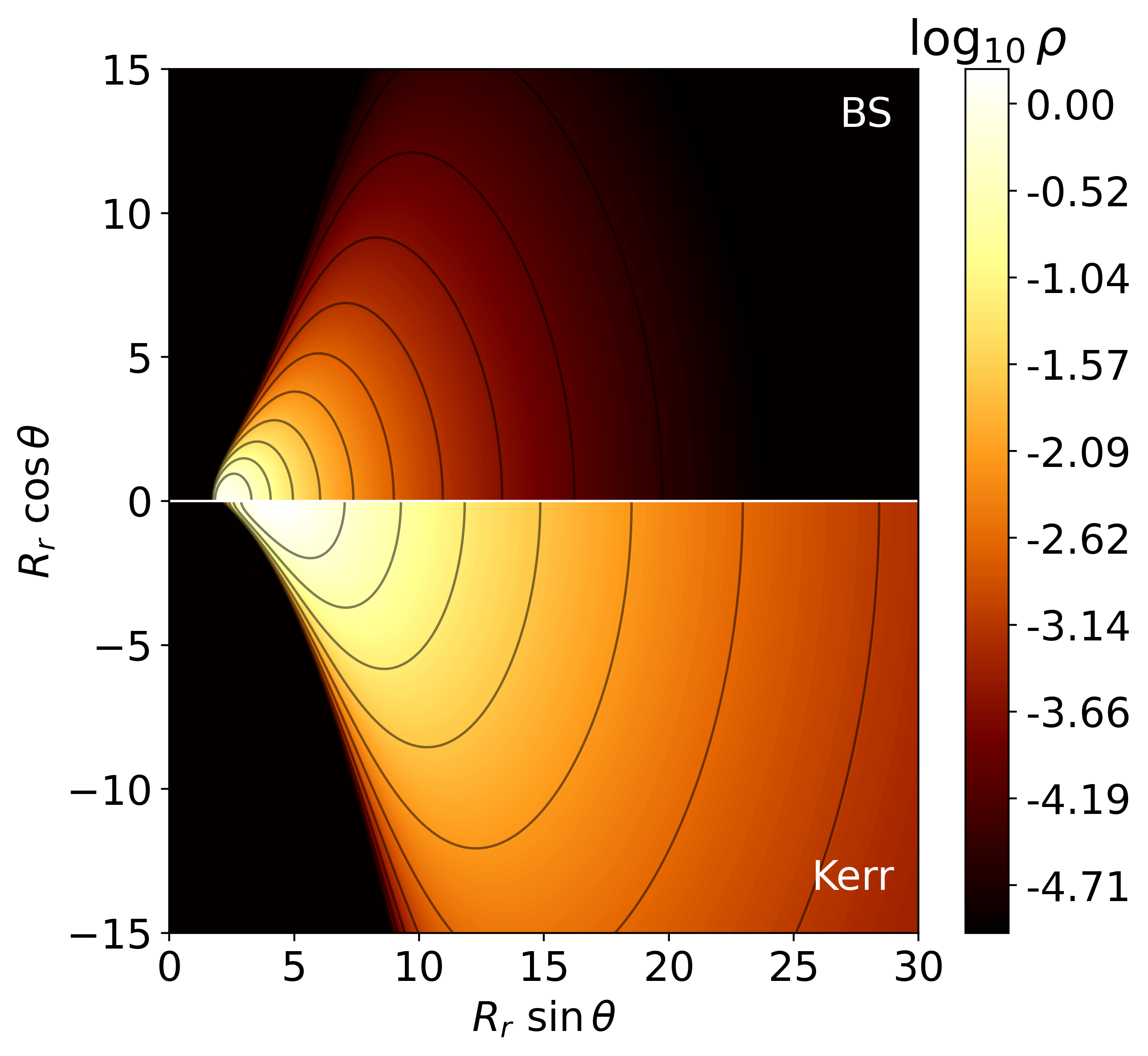}
  \caption{BS: 0.694, $\beta_{mc} = 10^{-5}$}
\end{subfigure}
\caption{Meridional cross section plots of the density distribution in pseudo-cylindrical coordinates. The upper half shows the BS torus, while the lower half shows the Kerr torus. The upper bound of the density scale is set to the maximum density between the BS and Kerr solution for the highest magnetization, $\log_{10}\rho_{\max} = 0.238$. The lower bound of the density scale is set to $\log_{10}\rho = -3$ for the BS 0.841 plots and to $\log_{10}\rho = -5$ for the BS 0.694 plots. Black curves mark equi-density surfaces.}
\label{fig:Torus_a825}
\end{figure}

The maximum density increases comparably for both BS solutions as well as for the Kerr solution. In contrast to the high and mid spin parameter solutions the equatorial density drops steeper for the BS disks for all degrees of magnetization. Thus, in this case we obtain more compact disks for the BS spacetimes than for the Kerr black hole. We can attribute this effect to the fact that the BS solutions become more relativistic when their angular frequency $\omega_{BS}$ decreases, which correlates with a decreasing spin parameter according to Fig.~$\ref{fig:BS_solutions}$. We see, that for the mildly relativistic BS with $\omega_{BS}= 0.841$ the disk density behaves in a very similar way to the Kerr disk. The BS 0.694 disk gets more compactified with increasing magnetization. The location of the density maxima is largely unaffected by magnetization for the BS disks, since the center location is already close to the inner edge for the non-magnetized disks. For the BS 0.841 disk the deviation of the location of the density maximum from the Kerr disk is small for all the degrees of magnetization ranging from $\approx 20 \%$ for non-magnetized disks to $\approx 10 \%$ for highly magnetized disks. It is more substantial for the BS 0.694 disks, where the values correspond to $\approx 40-50 \%$ (see also Fig.~\ref{fig:rel_diff}). Regarding the vertical density profile, the density curves of the BH and BS 0.694 disks behave more similarly for a high magnetization in contrast to the equatorial density profile. Regarding the vertical profile, the density of the BS 0.841 disks behaves similarly to the high and mid spin parameter solutions and drops less sharply, while the vertical density of the BS 0.694 solution behaves more similar to the BH disks, having a steep decline for all magnetization parameters. Fig.~\ref{fig:Torus_a825} presents the cross section plots, which showcase the greater extent of the Kerr disks compared to both BS disks through all degrees of magnetization. The radial extension $R_{out} - R_{\rho_{max}}$ of the non-magnetized BH disk is with $\approx 40 \%$ higher than the BS 0.841 disk as the difference decrease to $\approx 20 \%$ for high magnetization. For the BS 0.694 solutions the deviation of the disk extension from the Kerr BH is $\approx 80 \%$ irrespective of the magnetization.  As for the previous spin parameters that we considered, the angular velocity for the BSs depend extremely weakly on the magnetization, maintaining almost constant values. On the other hand, for the Kerr BH it increases, reaching almost 2 times higher values for highly magnetized disks compared to the non-magnetized case. As a result the highly magnetized BS 0.694 disks possess a similar rotation rate as the Kerr disks with a difference in the angular velocity ratio of $\approx~ 10\%$.

\begin{figure}[H]
\centering
\includegraphics[width=\linewidth]{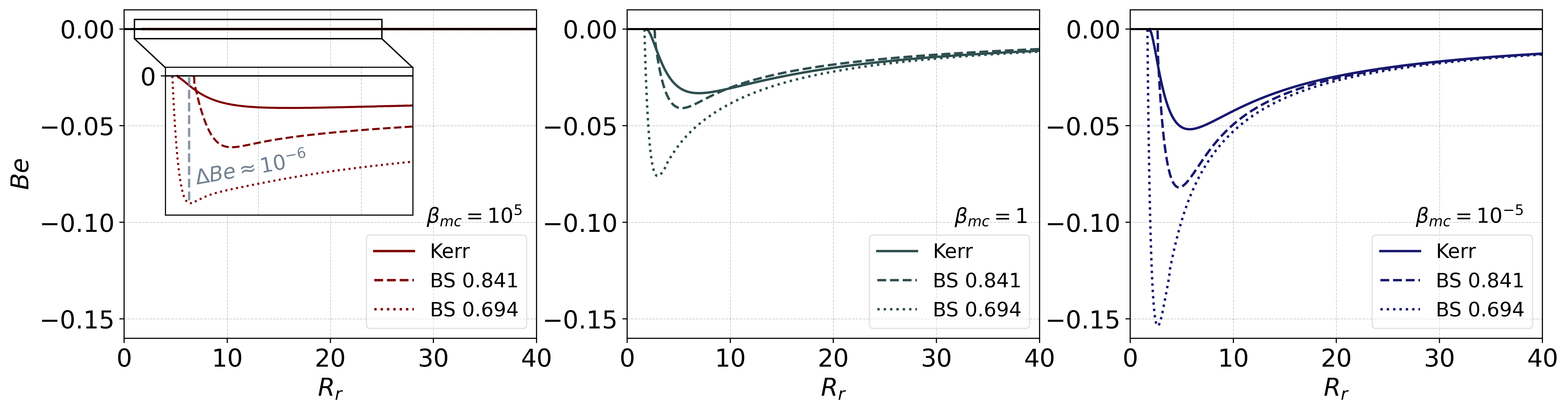}
\caption{Bernoulli parameter of the BS and Kerr disks in the equatorial plane for different degrees of magnetization, with the non-magnetized to the high magnetized solutions from left to right. For the non-magnetized disks a zoom-in plot is inserted for a better resolution, since $Be$ is nearly constant for both disks and close to $0$, with the maximum difference to 0 being on the order of $10^{-6}$.}
\label{fig:a825_Be}
\end{figure}

In accordance to the density distribution, the BS disks appear to be more energetically bound, as the Bernoulli parameter is more affected by magnetization compared to the BH disks (Fig.~\ref{fig:a825_Be}), for the strong magnetic fields the BS 0.694 disk has the lowest $Be$ peak. We conclude that in contrast to the high and mid spin parameter solutions, the BS disks are more affected by magnetization than the BH disk.

\begin{table}[H]
\centering
\vline
\begin{tabular}{c|c|c|c|c|c}
     \toprule
      $\beta_{mc} = 10^5$ & $\rho_{max}$ & $R_{\rho_{max}}$ & $R_{out} - R_{\rho_{max}}$ & $\frac{R_{in}}{R_{\rho_{max}}}$ & $\frac{\Omega_{\rho_{max}}}{\Omega_{in}}$ \\
     \midrule
     Kerr & 1 & 5.856 & 14.216 & 0.334 & 0.360 \\
     BS 0.841 & 1 & 4.796 & 7.786 & 0.549 & 0.373 \\
     BS 0.694 & 1 & 2.677 & 3.921 & 0.627 & 0.601\\
     \bottomrule
\end{tabular}
\vline
\\
\vline
\begin{tabular}{c|c|c|c|c|c|c}
     \toprule
      $\beta_{mc} = 1$ & $\rho_{max}$ & $R_{\rho_{max}}$ & $R_{out} - R_{\rho_{max}}$& $\Delta R_{rel}$ & $\frac{R_{in}}{R_{\rho_{max}}}$ & $\frac{\Omega_{\rho_{max}}}{\Omega_{in}}$\\
     \midrule
     Kerr & 1.134 & 4.855 & 7.737 & 0.171 & 0.403 & 0.460 \\
     BS 0.841 & 1.047 & 4.382 & 5.200 & 0.086 & 0.601  & 0.376 \\
     BS 0.694 & 1.049 & 2.492 & 2.533 & 0.069 & 0.674 & 0.609 \\
     \bottomrule
\end{tabular}
\vline
\\
\vline
\begin{tabular}{c|c|c|c|c|c|c}
     \toprule
     $\beta_{mc} = 10^{-5}$ & $\rho_{max}$ & $R_{\rho_{max}}$ & $R_{out} - R_{\rho_{max}}$ & $\Delta R_{rel}$ & $\frac{R_{in}}{R_{\rho_{max}}}$ & $\frac{\Omega_{\rho_{max}}}{\Omega_{in}}$\\
     \midrule
     Kerr & 1.634 & 4.214 & 5.439 & 0.280 & 0.464 & 0.545 \\
     BS 0.841 & 1.198 & 4.121 & 4.06 & 0.141 & 0.639 & 0.376 \\
     BS3 0.694 & 1.205 & 2.349 & 1.908 & 0.123 & 0.715 & 0.613 \\
     \bottomrule
\end{tabular}
\vline
\caption{Selected properties of the Kerr and BS disks in the equatorial plane, where $\rho_{max}$ is the maximum density, $R_{\rho_{max}}$ is the location of the density maximum,  $R_{out} - R_{\rho_{max}}$ is the equatorial distance from the density maximum to the effective outer edge of the disk, $\Delta R_{rel}$ is the relative difference in the location of the density maximum between magnetized and non-magnetized case, $\frac{R_{in}}{R_{\rho_{max}}}$ is the ratio between inner edge and density maximum location and $\frac{\Omega_{\rho_{max}}}{\Omega_{in}}$ is the ratio between the angular velocity at the density maximum and the inner edge of the disk.}
\label{tab3}
\end{table}

\subsection{Magnetization influence on the relative difference of disk quantities}

\begin{figure}[H]
\centering
\begin{subfigure}{.4\textwidth}
  \centering
  \includegraphics[width=\linewidth]{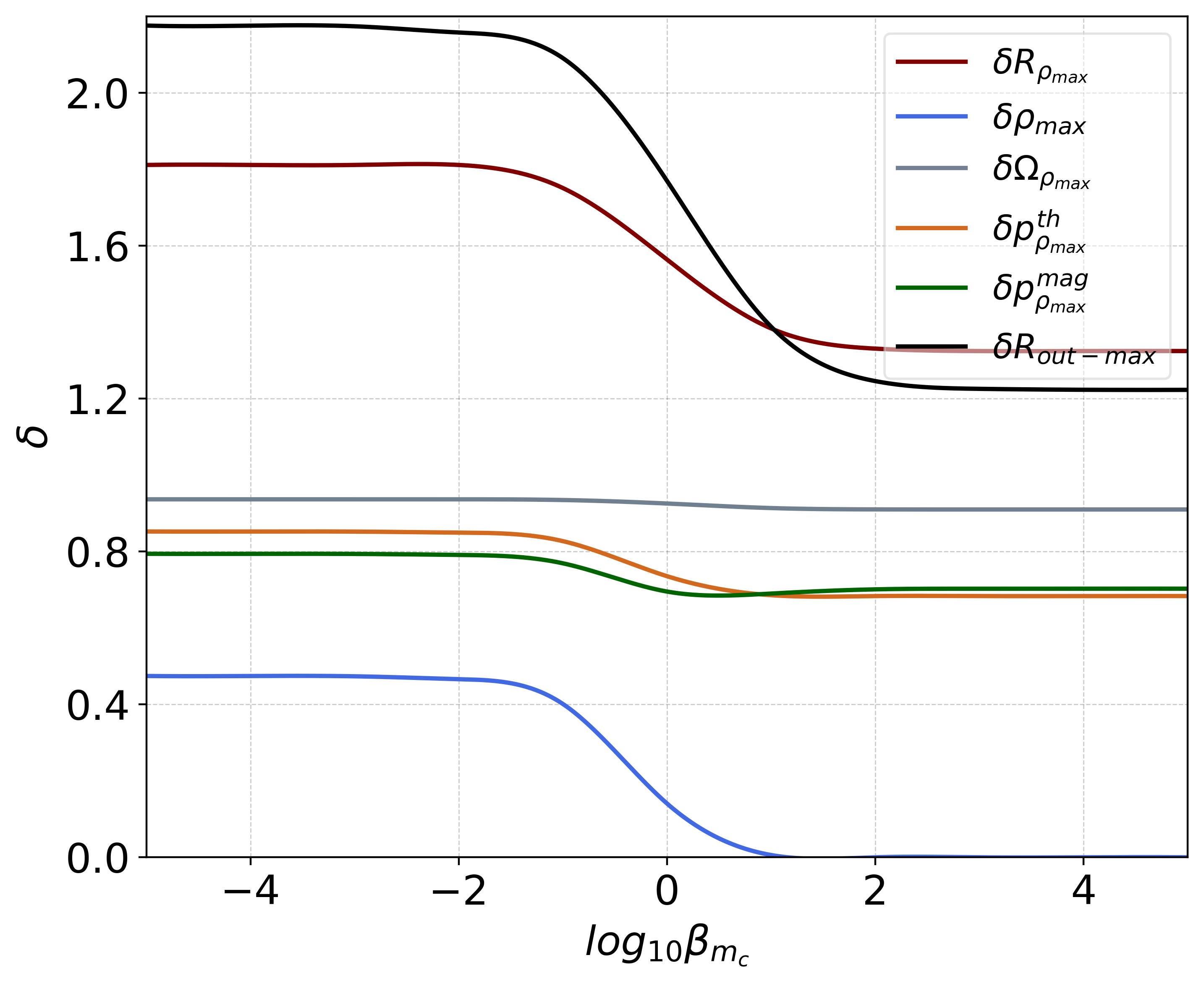}
  \caption{$a = 0.995$}
\end{subfigure}
\begin{subfigure}{.4\textwidth}
  \centering
  \includegraphics[width=\linewidth]{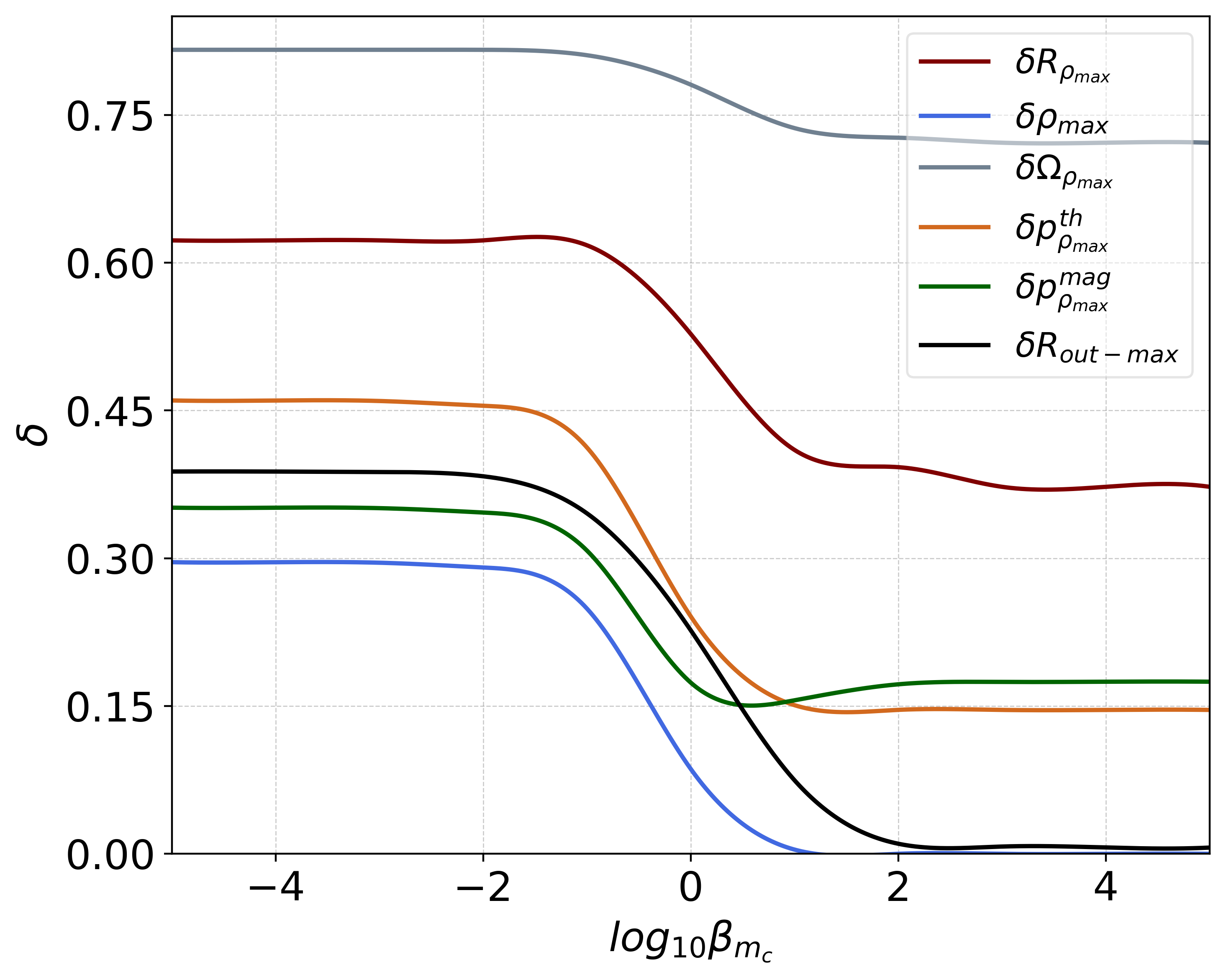}
  \caption{$a = 0.908$}
\end{subfigure}
\begin{subfigure}{.4\textwidth}
  \centering
  \includegraphics[width=\linewidth]{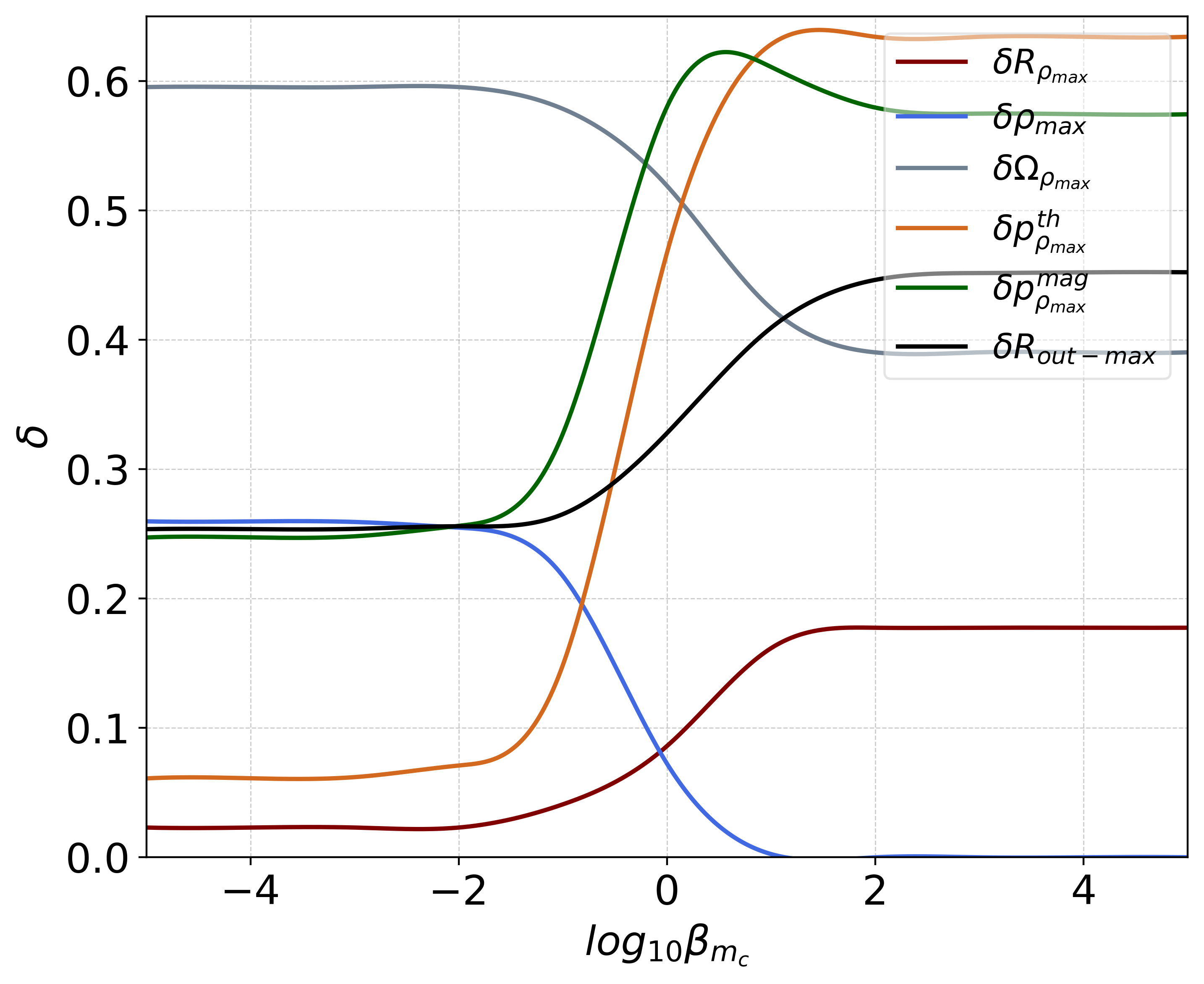}
  \caption{$a = 0.825$, $\omega_{BS} = 0.841$}
\end{subfigure}
\begin{subfigure}{.4\textwidth}
  \centering
  \includegraphics[width=\linewidth]{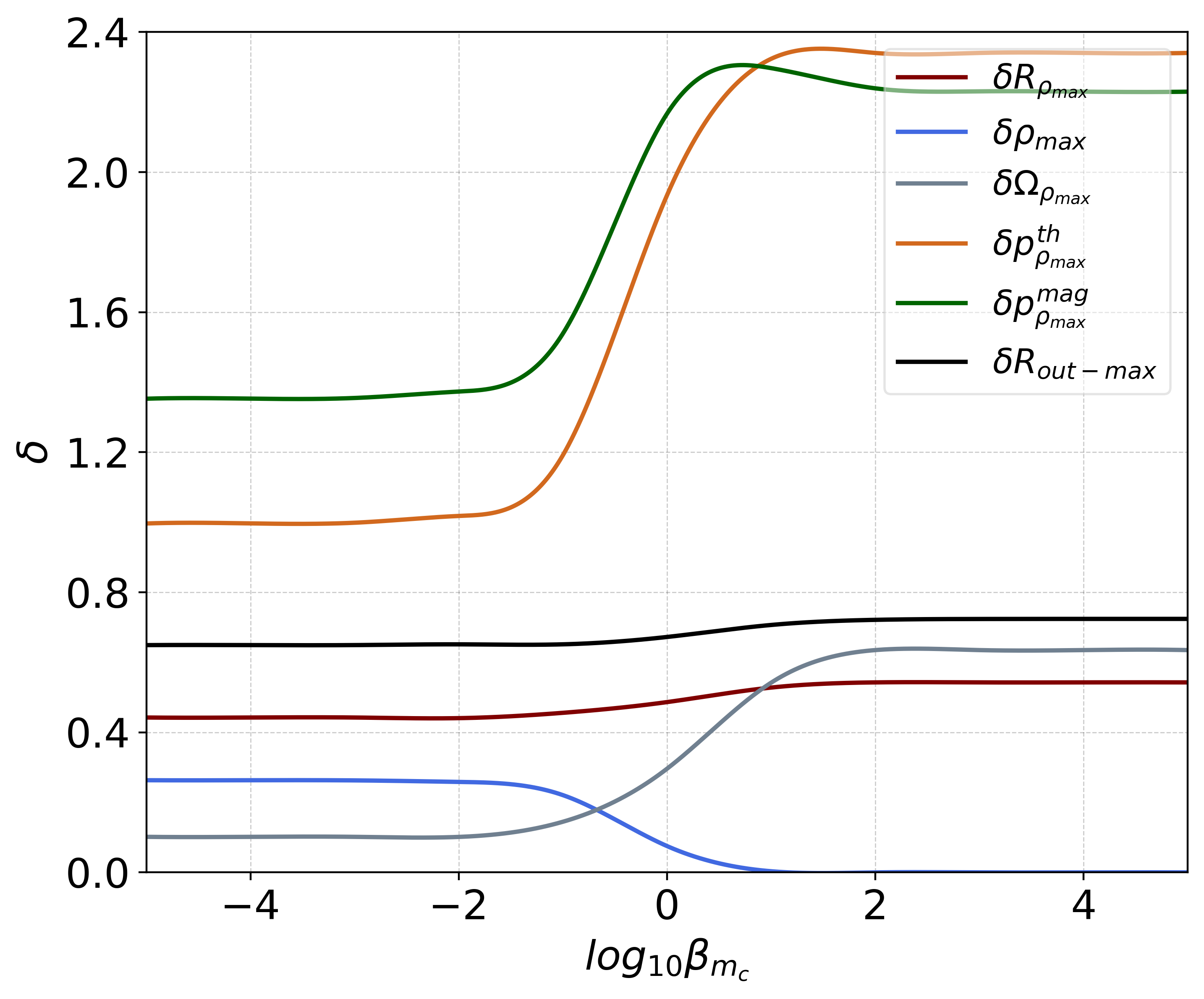}
  \caption{$a = 0.825$, $\omega_{BS} = 0.694$}
\end{subfigure}
\caption{Relative difference of selected disk quantities versus the magnetization parameter $\beta_{mc}$ for the different BS and Kerr solutions classified by their spin parameter. $\delta R_{\rho_{max}}$ is the relative difference of the density maximum location, $\delta \rho_{max}$ is the relative difference of the maximum density, $\delta \Omega_{\rho_{max}}$ is the relative difference of the angular velocity which a particle would have at the density maximum, $\delta p_{max}^{th}$ is the relative difference of the maximum thermodynamic pressure, $\delta p_{max}^{mag}$ is the relative difference of the maximum magnetic pressure and $R_{out - max}$ is the relative difference of the equatorial distance to the effective outer edge $R_{out}$. The highly magnetized disks are represented by a small value of the magnetization parameter and are therefore represented by the left side of the plots.}
\label{fig:rel_diff}
\end{figure}

The relative difference of a BS disk quantity $Q$ to its corresponding BH disk quantity $\Tilde{Q}$ can be expressed as $\delta Q = \frac{|\Tilde{Q}-Q|}{\Tilde{Q}}$. If a disk quantity is affected by the degree of magnetization, its relative difference can be expressed as a function of the magnetization parameter $\beta_{mc}$. The function $\delta Q(\beta_{mc})$ can therefore be used as a direct measure of the extent to which magnetic fields influence the similarities and differences between BS and Kerr disks. Fig.~\ref{fig:rel_diff} shows an analysis of the relative difference of selected disk quantities as a function of $\beta_{mc}$ for the previously shown high spin, mid spin and low spin solutions.

For the high and mid range spin parameter solutions the relative difference of every quantity is the highest for the strongly magnetized disks, as all quantities increase monotonically with a falling magnetization parameter. The lower spin parameter solutions showcase a different behavior, since the difference of the thermodynamic and magnetic pressure and also the difference of the density maximum location as well as the difference of the compactness parameter $R_{out - max}$ drop for highly magnetized disks. Regarding the BS 0.694 solution the only quantity difference which increases for highly magnetized disks is the density maximum. Overall the smallest differences in most of the quantities occur for the low spin and mildly relativistic BS 0.841 solution and the largest deviations for the high spin BS solution with $a = 0.995$.
We conclude that strong magnetic fields increase the difference in disk quantities for high and mid spin parameter solutions, whereas for low spin parameter solutions the opposite can occur for most of the general disk quantities. Strong magnetic fields could therefore lead to a correlation in accretion disk properties for some disk solutions in the lower spin parameter range.


\section{Retrograde Disks}
\begin{figure}[H]
\centering
\includegraphics[width=0.8\linewidth]{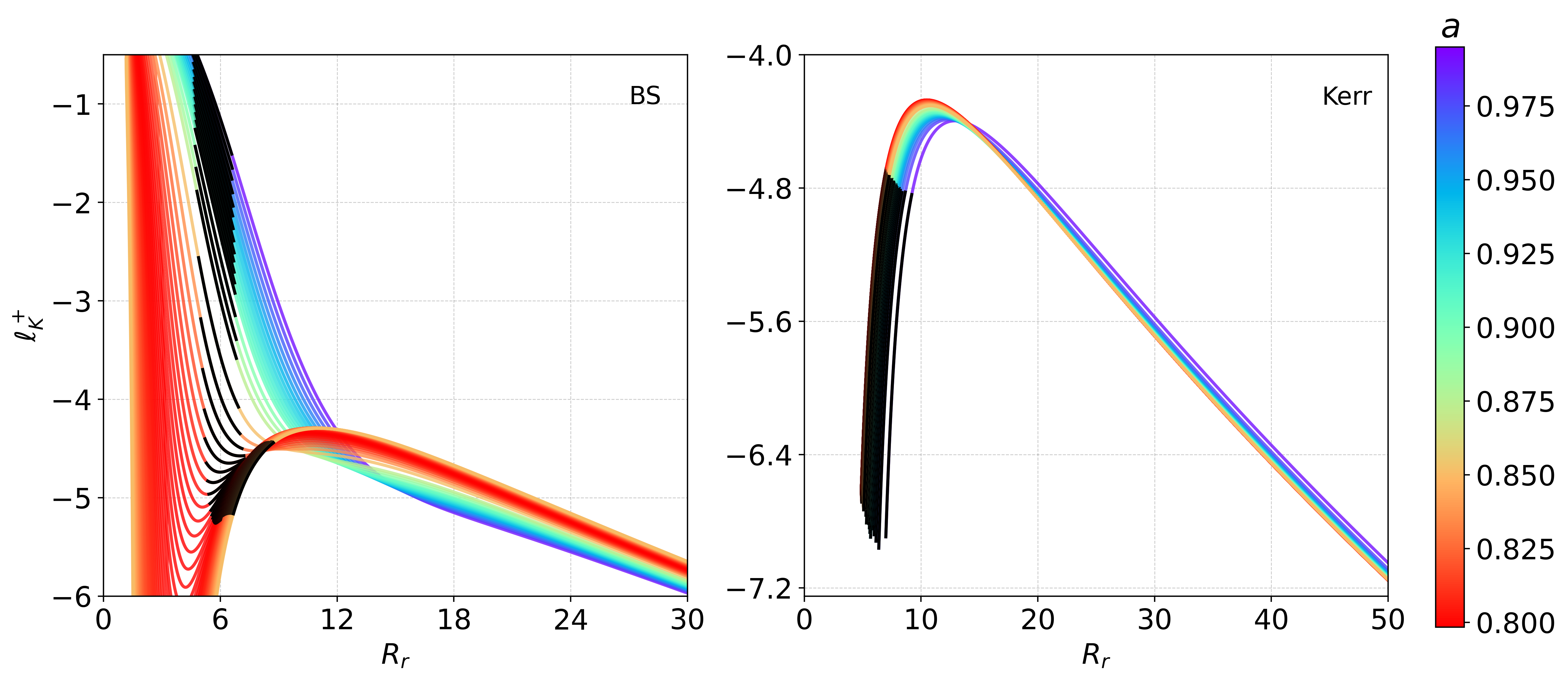}
\caption{Retrograde Keplerian specific angular momentum in the equatorial plane for the BS solutions on the left and Kerr solutions on the right. The BS and Kerr solutions are classified by their spin parameter $a$. Curve sections colored black mark intervals where only unbound orbits exist.}
\label{fig:BS_Kerr_ell-}
\end{figure}

In the case of retrograde disks, there are major differences between the possible BS and Kerr disk solution types.
The possible types for Kerr solutions are similar to the prograde disks, with cusp solutions possible for a lower absolute value of the Keplerian specific angular momentum and cuspless solutions for higher absolute values (Fig.~\ref{fig:BS_Kerr_ell-}).
The Keplerian specific angular momentum for BSs is non-monotonic for retrograde motion in contrast to prograde motion (Fig.~\ref{fig:BS_Kerr_ell-}). This imposes a variety of possible disk types, including two-centered solutions.

For high spin parameter BS solutions $\ell_K^-$ has a maximum and $\ell_0$ can have possibly three intersections with $\ell_K^-$. Depending on the second intersection these solutions mark either two-centered solutions, where the inner and outer center are connected by a cusp located at the second intersection, or a cuspless two-centered solution, where the disk is composed of an inner and outer torus. If the Keplerian specific angular momentum at the second intersection corresponds to an unbound orbit, the disk solution is cuspless. Under the influence of strong magnetic fields the density and pressure distributions of these kind of solutions can change to solutions with one density maximum, where the outer density maximum can get suppressed for magnetized disks. A more profound analysis can be found in \cite{Gjorgjieski}. Naturally there are more possible differences between retrograde disks for BSs and BHs, especially if the disk center is located close to the origin, since the angular momentum of the disk would then be of small absolute value. With increasing distance the specific angular momentum is increasing in absolute value monotonically for BHs as well as for BSs. For higher absolute values of the specific angular momentum there exist therefore cuspless one-centered solutions for BSs and BHs. These solutions would be located further away from the origin and they would be more comparable to each other in contrast to lower absolute angular momentum solutions, but astrophysical accretion disks are expected to have a specific angular momentum close to $\ell_{mb}$ and $\ell_{ms}$. In general, these solutions would therefore be of less interest than solutions where the disk center is located close to the origin (with a small absolute specific angular momentum of the disk). It might therefore be less difficult in the future to possibly distinguish retrograde BS disks from retrograde BH disks as compared to prograde disks.


\section{Conclusion}
We classified BS and Kerr solutions by a dimensionless spin parameter $a$ to compare their magnetized prograde thick disk solutions to each other. The influence of magnetization on the disk properties was isolated by fixing all other solution parameters besides the magnetization parameter $\beta_{mc}$. Disk solutions with the same spin parameter and the same specific angular momentum were compared to each other. For the specific angular momentum we assumed a uniform distribution throughout the disk and set its value to the Keplerian specific angular momentum value corresponding to the marginally bound orbit in the Kerr solution. The classification of BS and Kerr solutions defined a spin parameter range with its lower boundary given by the lowest spin parameter of the BS solutions, $a = 0.798$, and the upper boundary given by the Kerr limit $a = 1$. The lower half of the spin parameter range is composed of a BS solution possessing a twofold degeneracy, where two different BS solutions (classified by their angular frequency $\omega$) could have the same spin parameter. For an exemplary analysis of the whole spin parameter solution space, we picked solutions from the high, medium and low spin parameter range, namely $a = 0.995$, $a = 0.908$ and $a = 0.825$, whereby two BS solutions were corresponding to the $a = 0.825$ solution. Each of these exemplary solutions is capturing qualitatively the possible disk solution types of its represented spin parameter range. In order to undertake a qualitative comparison of accretion disks in different spacetime geometries we defined proper physical coordinates for the comparison, defined by the proper radial distance $R_r$. We computed and analyzed the properties of BS and BH disks for different degrees of magnetization, from non-magnetized disks ($\beta_{mc} \gg 1$) to mildly magnetized disks ($\beta \sim  1$) to highly magnetized disks ($\beta_{mc} \ll 1$) and compared them to each other. The density distribution acted as the main analyzed quantity, which was presented in the equatorial plane as well as for the vertical profile at the density maximum and also in a meridional cross-section plot. Besides other presented disk quantities the Bernoulli parameter was used as a further quantity to analyze the influence of magnetization on the disk stability. 

We found that magnetic fields strengthen the already present differences in non-magnetized disks for high spin parameter solutions such as the greater extent of the BS disks, especially in their vertical profile. The vertical thickness of the BS disks appears to be their main distinctive feature for the high spin parameter solutions, their properties are in general less affected by a high magnetization compared to the BH disks, which get more compactified with increasing magnetization. This is also reflected by the Bernoulli parameter, which is more affected by the magnetization for the BH disks. Therefore the high magnetized BS disks appear to be less energetically bound compared to the BH disk. In general, the properties of the high spin parameter BS disks are more resilient to changes in the degree of magnetization. 

Regarding the mid spin parameter solutions a similar behavior can be observed, with increasing magnetization the differences between BS and BH disks increase also. However, these differences are of less impact compared to the high spin parameter solutions, since the non-magnetized disks are already similar to each other and, furthermore, the disk solutions are affected with a comparable influence by magnetization (with a more compact Kerr solution again). Also the Bernoulli parameter is similarly influenced in both cases by $\beta_{mc}$, both disks are energetically similarly bound. Nevertheless, like for the high spin parameter solutions, the greater vertical thickness can be identified as a main distinctive feature of the BS disks compared to the Kerr disks.

In case of the lower spin parameter an opposite behavior compared to the high and mid spin parameter solutions can be observed.
With increasing magnetization the differences between BS and BH disks decrease. The equatorial extent of the BH disk is now greater compared to both BS disks for all degrees of magnetization, but as in the high and mid spin parameter solutions, the vertical density for the BH disk is stronger affected by magnetization. All three disk solutions are compactified similarly, with the highly relativistic (lower angular frequency) BS solution being the most influenced by strong magnetic fields.

We summarize, BS and BH disks get both compactified with increasing magnetization and tend to be more bound, since their equatorial Bernoulli parameter is developing a clear minimum for the highly magnetized disks. The differences in the main disk properties between BS and BH disks may increase for highly magnetized disks or even decrease, depending on the spin parameter of the BS and Kerr solutions. For high and mid range spin parameter solutions magnetic fields strengthen the differences and for low spin parameter solutions the differences may decrease, depending on the BS solution. Strong magnetic fields could lead therefore to a correlation in disk properties of accretion disks around different types of compact objects, such as BS and BHs, as shown. Our results demonstrate that highly magnetized disks around fast rotating boson stars deviate most strongly from the Kerr black hole. Thus, astrophysical systems with the corresponding properties could be a suitable laboratory for detecting boson stars. On the other hand, highly magnetized disks around slowly rotating mildly relativistic boson stars resemble closely the corresponding disks around Kerr black holes. Therefore, a boson star with the described physical conditions could represent a good Kerr black hole mimicker. We identify the vertical profile of BS disks as the main distinctive feature compared to BH disks for high and mid spin parameter solutions and their smaller equatorial extent for low spin parameter solutions.

\begin{acknowledgments}
We would like to thank Lucas Collodel and Matheus Teodoro for discussions and BS data. JK gratefully acknowledges support by DFG project Ku612/18-1. PN is partially supported by the Bulgarian NSF Grants KP-06-H38/2 and KP-06-H68/7.
\end{acknowledgments}    

\bibliography{literature}

\end{document}